\documentclass[longauth]{aa}  

\usepackage{graphicx}
\usepackage{txfonts}
\usepackage{lipsum}
\usepackage{subcaption}         
\usepackage{lscape}             
\usepackage{placeins}           
\usepackage{wrapfig}
\usepackage{floatrow}

\usepackage[colorlinks=true,linkcolor=blue,citecolor=blue,urlcolor=blue]{hyperref}%
\usepackage[dvipsnames]{xcolor}
\usepackage{upgreek}

\newcommand\hst{{HST}\xspace}
\newcommand\jwst{{JWST}\xspace}
\newcommand\bagpipes{{\sc Bagpipes}}
\newcommand\msaexp{{\sc Msaexp}}
\newcommand\grizli{{\sc Grizli}}
\newcommand\eazy{{\sc Eazy}}
\newcommand\cloudy{{\sc Cloudy}}
\newcommand\av{$A_V$}
\newcommand\sfrt{${\rm SFR_{10\ Myr}}$}
\newcommand\sfrh{${\rm SFR_{100\ Myr}}$}

\newcommand\ugis{$(ugi)_s$}
\newcommand\us{$u_s$}
\newcommand\gs{$g_s$}
\newcommand\is{$i_s$}
\newcommand\ug{$(u_s-g_s)_0$}
\newcommand\gi{$(g_s-i_s)_0$}

\usepackage{seqsplit}

\begin{document}

\title{The DAWN \jwst Archive:\\ A spectroscopically-informed population study of the physical properties of galaxies at \texorpdfstring{$z>2$}{Lg}}

\titlerunning{A spectroscopically-informed population study of the physical properties of galaxies at \texorpdfstring{$z>2$}{Lg}}
\authorrunning{Gottumukkala et al.}

\author{Rashmi~Gottumukkala\inst{1,2}\thanks{Corresponding author; {\tt rashmi.gottumukkala@gmail.com}}
   \and Gabriel~Brammer\inst{1,2}
   \and Sune~Toft\inst{1,2}
   \and Mengyuan~Xiao\inst{3}
   \and Anna~de~Graaff\inst{4,5}\thanks{Clay Fellow}
   \and Pascal~A.~Oesch\inst{3,1,2}
   \and Francesco~Valentino\inst{1,6}
   \and Katherine~E.~Whitaker\inst{7,1,2}
   \and William~M.~Baker\inst{8}
   \and Laia~Barrufet\inst{9}
   \and Rachel~Bezanson\inst{10}
   \and Leindert~A.~Boogaard\inst{11}
   \and Alex~J.~Cameron\inst{1,2}
   \and Olivia~R.~Cooper\inst{12}\thanks{NSF Astronomy and Astrophysics Postdoctoral Fellow}
   \and Emma~Giovinazzo\inst{3}
   \and Kasper~E.~Heintz\inst{6,1,2}
   \and Michaela~Hirschmann\inst{13}
   \and Kei~Ito\inst{1,6}
   \and Ivo~Labbe\inst{14}
   \and Minju~M.~Lee\inst{1,6}
   \and Zach~J.~Lewis\inst{15}
   \and Jorryt~Matthee\inst{16}
   \and Ian~McConachie\inst{15}
   \and Desika~Narayanan\inst{17,1,2}
   \and Clara~L.~Pollock\inst{1,2}
   \and Sedona~H.~Price\inst{18}
   \and David~J.~Setton\inst{19}\thanks{NHFP Hubble Fellow}
   \and Marko~Shuntov\inst{1,2}
   \and Pieter van Dokkum\inst{20}
   \and Bingjie~Wang\inst{21}\thanks{NHFP Hubble Fellow}
   \and Joris~Witstok\inst{1,2}
   \and Yunchong~Zhang\inst{10}
   \and Pengpei~Zhu\inst{1,6,22}
   }

\institute{Cosmic Dawn Center (DAWN), Copenhagen, Denmark
   \and Niels Bohr Institute, University of Copenhagen, Jagtvej 128, DK-2200 Copenhagen N, Denmark
   \and Department of Astronomy, University of Geneva, Chemin Pegasi 51, 1290 Versoix, Switzerland
   \and Max-Planck-Institut f\"ur Astronomie, K\"onigstuhl 17, D-69117 Heidelberg, Germany
   \and Center for Astrophysics $|$ Harvard \& Smithsonian, 60 Garden St., Cambridge MA 02138, USA
   \and DTU Space, Technical University of Denmark, Elektrovej 327, DK2800 Kgs. Lyngby, Denmark
   \and Department of Astronomy, University of Massachusetts Amherst, Amherst MA 01003, USA
   \and DARK, Niels Bohr Institute, University of Copenhagen, Jagtvej 155A, DK-2200 Copenhagen, Denmark
   \and Institute for Astronomy, University of Edinburgh, Royal Observatory, Edinburgh, EH9 3HJ
   \and Department of Physics and Astronomy and PITT PACC, University of Pittsburgh, Pittsburgh, PA 15260, USA
   \and Leiden Observatory, Leiden University, PO Box 9513, NL-2300 RA Leiden, The Netherlands
   \and Department for Astrophysical \& Planetary Science, University of Colorado, Boulder, CO 80309, USA
   \and Institute of Physics, Laboratory for Galaxy Evolution, EPFL, Observatoire de Sauverny, Chemin Pegasi 51, CH-1290 Versoix, Switzerland
   \and Centre for Astrophysics and Supercomputing, Swinburne University of Technology, Melbourne, VIC 3122, Australia
   \and Department of Astronomy, University of Wisconsin-Madison, 475 N. Charter St., Madison, WI 53706, USA
   \and Institute of Science and Technology Austria (ISTA), Am Campus 1, 3400 Klosterneuburg, Austria
   \and Department of Astronomy, University of Florida, 211 Bryant Space Sciences Center, Gainesville, FL 32611, USA
   \and Space Telescope Science Institute, 3700 San Martin Drive, Baltimore, MD 21218, USA
   \and The William H. Miller III Department of Physics and Astronomy, The Johns Hopkins University, Baltimore, MD 21218
   \and Department of Astronomy, Yale University, New Haven, CT 06520, USA
   \and Department of Astrophysical Sciences, Princeton University, Princeton, NJ 08544, USA
   \and INAF-Osservatorio Astrofisico di Arcetri, Largo Enrico Fermi 5, I-50125 Firenze, Italy
   }

   \date{Received \today}

\abstract{
Charting the evolution of star-formation, stellar mass and dust build-up over cosmic time relies on the accurate characterisation of the stellar population properties of high redshift galaxies.
Here, we present a catalogue of spectroscopically derived physical properties for $\sim 12,000$ galaxies, with archival \jwst{}/NIRSpec PRISM spectroscopy and \jwst{}/NIRCam imaging. Our sample includes all publicly available high quality PRISM spectra at $z_{\rm spec}>2$, uniformly reduced and available on the DAWN \jwst Archive (DJA). We present a custom version of the spectral energy distribution (SED) fitting code \bagpipes{}, used to perform joint spectrophotometric SED modelling of PRISM spectra and available broadband NIRCam photometry. In doing so, we publicly release a science-ready catalogue characterising the physical properties of galaxies at cosmic noon and beyond. We report key physical properties, such as stellar masses, star-formation rates, star-formation timescales, and V-band dust attenuation (\av{}).
We further probe the recent star-formation histories (SFH) of galaxies via their star-formation rates on 3 Myr, 10 Myr, and 100 Myr timescales, capturing both rising and falling recent SFHs.
Through dust-corrected \ugis{} colour-colour diagrams of massive galaxies, we show how quiescent and dusty galaxies populate this colour space at $z>2$, with galaxies experiencing a recent downturn in star-formation occupying distinct parameter space.
We additionally study the dustiest galaxies in our sample, showing that they are preferentially more massive and strongly star-forming relative to the parent sample, residing on the high-mass end of the star-forming main-sequence. The slope of the dust attenuation curve has a strong dependence on \av{}, varying from steep (higher dust index) to shallow (lower dust index) as the dust attenuation increases and the specific star-formation rate decreases.
Our findings show that the unparalleled number statistics of this catalogue of spectroscopically derived stellar population properties uniquely enables the study star-formation, quenching and dust properties of galaxies from cosmic noon to cosmic dawn.

}


\maketitle

\nolinenumbers

\section{Introduction}
\label{sec:intro}

Characterising the properties of the stellar populations of high-redshift galaxies is central to our understanding of how galaxies evolve through cosmic time \citep{Tinsley1980}. Determining fundamental properties such as the stellar masses ($M_\star$), star-formation rates (SFR) and dust attenuation (\av{}) of galaxies together with an accurate reconstruction of their star-formation histories (SFHs) up to the highest observable redshifts is key to informing us about the nature of star-formation in the high-redshift universe \citep{ForsterSchreiber2020,Stark2016}, the build-up of stellar mass over cosmic time \citep{Madau2014}, and the emergence of the first dusty \citep[e.g.][]{Casey2014, Wang2019} and quiescent galaxies \citep[see][for a recent review]{Whitaker2026}.

Accurately mapping empirical observations to the physical properties of galaxies requires detailed spectral energy distribution (SED) modelling of the rest-frame ultraviolet (UV), optical, and near-infrared (NIR) emission of galaxies \citep[e.g.,][]{Walcher2011, Conroy2013}. The rest-frame UV and optical, together, are most sensitive to the youngest stellar populations and their surrounding H\,{\sc ii} regions, and provide constraints on the most recent epochs of star-formation \citep{Kennicutt1998-review, Kennicutt2012}. The rest-frame optical and NIR instead enable the characterization of older stellar populations \citep{Kauffmann2003}, the effects of dust attenuation in the interstellar medium (ISM), and constrain the SFHs of galaxies over longer timescales \citep{Conroy2013}. Obtaining dense rest-frame wavelength coverage from the UV to the NIR is central to accurately constraining the SFHs of galaxies through cosmic time \citep[see e.g.][for an overview on measuring SFHs]{Carnall2019b,Leja2019}.

Deep photometric surveys in the UV and optical were pioneered using the Hubble Space Telescope (\hst), beginning with the Hubble Deep Field \citep[HDF;][]{Williams1996} and subsequently expanding to the Cosmic Assembly Near-infrared Deep Extragalactic Legacy Survey (CANDELS) fields \citep{Grogin2011, Koekemoer2011}. Large photometric studies from such surveys (often complemented by NIR data from ground-based observatories such as the Keck Observatory and the Very Large Telescope, as well as data from the Spitzer Space Telescope) enabled the establishment of the star-forming main sequence relation \citep{Noeske2007,Elbaz2007, Whitaker2012_ms, Speagle2014, Renzini2015}, the evolution of cosmic SFRD \citep{Madau2014}, the build-up of stellar mass across cosmic time \citep{Stefanon2015, McLeod2021, Weaver2022a} and the emergence of high-redshift `red and dead' galaxies \citep{Franx2003, ForsterSchreiber2004, Glazebrook2017, Valentino2020}. The relatively shallow and lower resolution rest-NIR imaging from Spitzer was upgraded with the launch of James Webb Space Telescope \citep[\jwst;][]{Gardner2023}, which has transformed our view of the high-redshift universe by providing deep, high-resolution NIR imaging surveys \citep[e.g.][]{Eisenstein2025, Finkelstein2023, Bezanson2022, Casey2023, Dunlop2021}, perfectly complementing the wavelength coverage and spatial resolution of \hst such that the rest-frame UV and optical emission of galaxies can now be studied up to the highest redshifts. New constraints push our understanding of stellar mass build-up \citep{Weibel2024_smf, Shuntov2025, Harvey2025_EPOCHS}, the nature of star formation in the early universe \citep{Tacchella2022, Simmonds2025, Looser2025}, and the emergence of the first dusty and quiescent galaxies \citep{Barrufet2023, Gonzalez2023, Rodighiero2023, Valentino2023, Gottumukkala2024, Manning2025, Baker2025, Barrufet2026} to even earlier epochs.

However, photometric surveys are plagued by inherent limitations. Foremost, uncertainties in photometric redshifts can have cascading effects on the determination of galaxy physical properties by enhancing degeneracies between age, dust, and mass. For example, massive dusty galaxies can masquerade as ultra-high-redshift objects \citep{Naidu2022, Zavala2023, ArrabalHaro2023_pub}, little red dots \citep{Matthee2023} can have optical colours similar to those of bona fide early massive galaxies \citep{Labbe2023a}, and the Balmer and 4000 \AA{} breaks can mimic dust-reddened stellar continua \citep{Wilkins2024}. In the case of Lyman-break galaxies, photometric redshifts can be significantly overestimated in the presence of damped Ly$\alpha$ absorption \citep[DLA;][]{Heintz2024_DLA, Heintz2026}. While the inclusion of medium-band photometry in surveys can make significant headway in improving photometric redshift estimates \citep[see e.g.][]{Whitaker2011, Williams2023_jems, Suess2024_megascience, Muzzin2025}, it nevertheless remains a challenge to accurately determine properties such as star-formation histories, which are encoded as detailed features in the stellar continuum of galaxies.

Spectroscopic surveys aim to alleviate these tensions. Large-scale spectroscopic surveys like 3D-HST enabled the accurate characterisation of the stellar population properties of galaxies at $1<z<3.5$ \citep{vanDokkum2011, Brammer2012, Momcheva2016}. In contrast, spectroscopic surveys with ground-based observatories, such as LEGA-C \citep{vanderWel2021}, VANDELS \citep{Pentericci2018, McLure2018} and KMOS$^{\rm 3D}$ \citep{Wisnioski2019} with the Very Large Telescope and Keck/MOSFIRE \citep{Schreiber2015, Belli2019}, provided detailed constraints on the stellar populations of massive galaxies \citep[e.g.][]{Kaushal2024} and quiescent galaxies up to $z\sim4$ \citep{Schreiber2018, Valentino2020}.  However, these surveys were fundamentally limited beyond $z\sim4$, where rest-frame optical diagnostics shift out of ground-based NIR coverage and only the brightest, most massive galaxies remained accessible.

The Near Infrared Spectrograph \citep[NIRSpec;][]{Jakobsen2022} on \jwst has truly revolutionised the large-scale study of high-redshift galaxies with detailed spectroscopy at $z>2$. With its Multi Object Spectrograph \citep[MOS;][]{Ferruit2022} capable of taking hundreds of spectra simultaneously in a single pointing, and with continuous wavelength coverage from ${\rm [0.5, 5.5]\ \mu m}$ with the NIRSpec/PRISM mode, NIRSpec provides access to the full rest-frame optical window from $z\sim0.5$ to $z\sim7$, predominantly limited to the rest-frame UV beyond this. Numerous spectroscopic surveys \citep[see e.g.][amongst several others]{Fujimoto2023, Maseda2024, Price2024_uncover_spectra, Dickinson2024, deGraaff2025_RUBIES, Eisenstein2025, Naidu2026}
have helped push the observable redshift frontier to $z>14$ \citep{Carniani2024, Naidu2026} and enabled the study of the UV-bright high-redshift universe in exquisite detail \citep[e.g.][]{Curtis-Lake2023, Witstok2025, Cameron2023, Roberts-Borsani2024, Heintz2025}. In addition, other spectroscopic surveys target $z>3$ massive dusty galaxies \citep{Barrufet2024, Barrufet2024_jwstprop} and massive quiescent galaxies \citep{Glazebrook2024, Carnall2024, Ito2025, Slob2024} for the first time, allowing for the spectroscopically-informed characterisation of the stellar population properties and SFHs of galaxies at $z>4$ \citep[e.g.][]{Nanayakkara2024, Nanayakkara2024b, Carnall2023, deGraaff2024a, Baker2024_quiescent_SFHs}.

With numerous \jwst imaging and spectroscopic surveys contributing to an ever-growing archive of public data, the DAWN \jwst Archive (DJA)\footnote{\protect\url{https://dawn-cph.github.io/dja/}} \citep{Brammer2025_DJA-spec} is an initiative aimed at homogeneously reducing these data and providing science-ready data products for timely use by the extragalactic astronomical community. The DJA currently provides reduced v7 NIRCam imaging data \citep{Valentino2023} for all extragalactic legacy fields and $>80,000$ NIRSpec spectra (reduced v4.5) across the low, medium, and high resolution gratings, which have all been visually inspected and graded \citep{deGraaff2025_RUBIES, Heintz2025, Valentino2025, Pollock2026}. Studies using the DJA as a whole have shown that by combining all public imaging and/or spectroscopic \jwst datasets, the resulting number statistics give unique insights into the nature of galaxies in the high-redshift universe \citep[e.g.][amongst several other works]{Ito2025, Heintz2025, Pollock2026, Genin2025, Lee2025, deGraaff2025_LRDoverview}.

In this paper, we present the first version of a stellar population synthesis modelling catalogue on the DJA, providing spectral energy distribution (SED) fits, SFHs, and posterior physical parameters of $\sim 12,000$ galaxies. These galaxies are selected from \jwst observations with high quality NIRSpec/PRISM spectra ({\tt grade=3}) and wide-band NIRCam photometry, and are homogeneously modelled with a custom version of \bagpipes{} \citep{Carnall2018, Carnall2019}. Our dataset spans the period of time from the peak epoch of star formation ($z=2$) to our current frontier at cosmic dawn \citep[$z=14.4$;][]{Naidu2026}. We present an overview of the physical properties and SFHs of this sample, and show that it spans an incredibly diverse range of SED shapes, SFHs, and stellar ages. We demonstrate that our spectrophotometric fitting is sensitive to SFRs on various timescales, allowing us to robustly constrain
the recent SFH. We can track the spectroscopically-derived physical properties of a large sample of massive galaxies across colour-colour space, and are sensitive to the detailed dust properties of galaxies beyond cosmic noon. An application of this dataset can be found in  \cite{Xiao2026}, where we trace how the properties of the most massive spectroscopically observed galaxies evolve across cosmic history.

This paper is structured as follows. In Section \ref{sec:data}, we present the DJA spectroscopic and photometric database, including an overview of the data reduction. The details of our spectroscopic sample selection from the DJA, and a discussion of the caveats of selecting such a sample from a heterogeneous set of extragalactic \jwst surveys, can also be found in this section. In Section \ref{sec:SED_modelling}, we present our SED modelling framework with \bagpipes{} \citep{Carnall2018, Carnall2019}, introducing a modified version of the code that adds line marginalisation flexibility and an evaluation of its efficacy.
In Section \ref{sec:results}, we present an overview of our catalogue of SED-modelled properties, including the physical properties, multi-timescale SFRs, and recent SFHs of massive galaxies, their behaviour in \ugis{} colour-colour space, and their dust attenuation characteristics. We discuss and conclude our work in Sections \ref{sec:discussion} and \ref{sec:conclusion}.

For this work, we assume a flat $\Lambda$CDM cosmological model with H$_0 = 67.7 \ \text{km}\ \text{s}^{-1}\ \text{Mpc}^{-1}$ and $\Omega_m = 0.3$ and $\Omega_\Lambda = 0.7$ as found by the \cite{Planck2014}. All magnitudes are quoted in the AB system \citep{Oke1983}.

\section{Data and sample selection}
\label{sec:data}

The following sections describe the imaging and spectroscopic data products used in this paper, which are available on the DAWN \jwst Archive (DJA) \citep{Brammer2025_DJA-spec}, a public repository of \jwst/NIRCam and \jwst/NIRSpec extragalactic survey science-ready data products, complemented by ancillary \hst/ACS and \hst/WFC3 data from several \hst surveys \citep[including CANDELS;][]{Grogin2011, Koekemoer2011, Stefanon2017b}. The \jwst imaging and spectroscopic data are reduced with the \grizli{}\footnote{\protect\url{https://github.com/gbrammer/grizli}} and \msaexp{}\footnote{\protect\url{https://github.com/gbrammer/msaexp}} software, as outlined below.

\subsection{Imaging data, photometry and catalogues}
\label{subsec:imaging_data_and_photometry}

The level-2 imaging data products were retrieved from the Mikulski Archive for Space Telescopes (MAST) and were reduced with the \grizli{} pipeline \citep{Brammer2023}. This reduction includes updated NIRCam zeropoints and sky flats. The images were aligned to Gaia DR3 \citep{Gaia2021}.

Sources are extracted from an optimally noise-weighted, combined detection image that is created from three NIRCam Long Wavelength filters (F277W, F356W and F444W). The photometry is then extracted in circular apertures of 0.5'' and is scaled to values within an elliptical Kron aperture \citep[see][]{Skelton2014}. While the photometry are not PSF matched, aperture (``zero-point'') corrections are applied to all bands based on calculations done on the detection image. The imaging data reduction is described and further detailed in \cite{Valentino2023} (see also references therein).

All sources are SED-modelled with the \eazy{}\footnote{\protect \url{https://github.com/gbrammer/eazy-py}} software \citep{Brammer2008}, in order to estimate photometric redshifts and rest-frame colours. The modelling details can be found in \cite{Valentino2023} and the data products are available on the DJA.

For this work, we use data from the following \jwst/NIRCam broadband filters: F090W, F115W, F150W, F200W, F277W, F356W, and F444W. For each source, we use the maximum number of available wide bands, requiring a minimum of 2 bands for photometric calibration of the spectrum (see Section \ref{subsec:phot_calibration}). $>95\%$ of sources include either 6 or 7 photometric bands. While photometry from additional \hst/ACS and \hst/WFC3 bands is available on the DJA for several of the survey fields, for uniformity we use only the \jwst/NIRCam wide bands in this work. A summary of extragalactic fields and photometric catalogues is provided in Table \ref{tab:phot-cats} in the Appendix.

\begin{figure*}[t!]
\includegraphics[width=\linewidth]{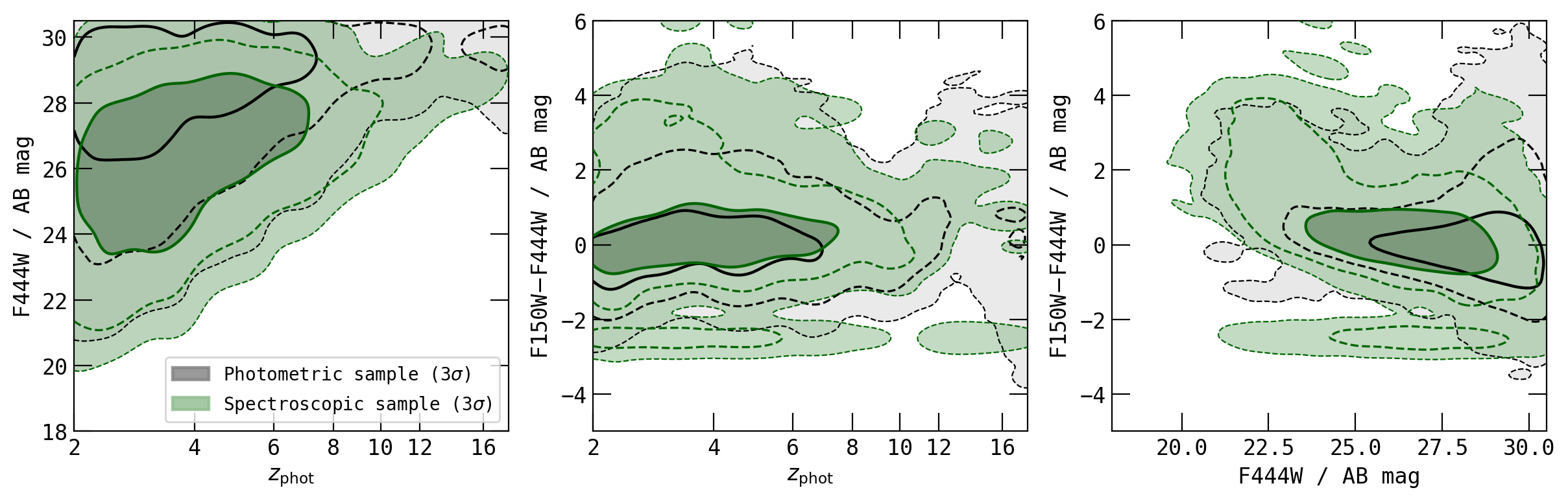}
\caption{AB magnitude in the F444W photometric band vs. photometric redshift (left), F150W-F444W colour vs. photometric redshift (centre), and F150W-F444W colour vs. F444W magnitude. In each panel, the black contours represent the $1\sigma$ (solid line) $2\sigma$ and $3\sigma$ (dashed lines) levels of the photometric sample and the green contours show the same for the DJA {\tt grade==3} PRISM spectroscopic sample.}
\label{fig:sample_colour_space_0}
\end{figure*}

\subsection{Spectroscopic data and reduction}
\label{subsec:spectroscopic_data_and_reduction}

For this study, we use data from the NIRSpec Multi-Object Spectroscopy (MOS) observing mode with the micro-shutter array \citep[MSA][]{Jakobsen2022, Ferruit2022}. The MSA is composed of $\sim 250,000$ shutters, each 0.20''$\times$0.46'' in area.
With a 3.6'$\times$3.4' field of view, the MSA can obtain hundreds of spectra simultaneously. NIRSpec has several disperser-filter combinations with varying levels of resolution. For this work, we use data from the PRISM/CLEAR combination, which has a nominal resolving power of $\sim 30-300$ across a wavelength range of 0.5 to 5.5 ${\rm \mu m}$.

We use the latest v4.5 reductions of the NIRSpec PRISM spectra on the DJA,
reduced and processed with the \msaexp{} software \citep{Brammer2025_DJA-spec,brammer_2023_msaexp}. Data reduction and processing are described in \cite{deGraaff2025_RUBIES, Heintz2025, Valentino2025,Pollock2026}. Data products, including the reduced spectra and post-processed properties, such as emission line fluxes, equivalent widths (EWs) and spectroscopically derived rest-frame colours, are available on the DJA.

The PRISM data used in this work are drawn from several \jwst extragalactic surveys, all publicly available on MAST. Spectroscopic surveys include the Canadian NIRISS Unbiased Cluster Survey \citep[CANUCS;][]{Willott2022}, the CANDELS-Area Prism Epoch of Reionization Survey \citep[CAPERS][]{Dickinson2024}, the Cosmic Evolution Early Release Science Survey \citep[CEERS][]{ArrabalHaro23, Fujimoto2023}, the NIRSpec WIDE GTO Survey \citep[WIDE][]{Maseda2024}, the JWST Advanced Deep Extragalactic Survey \citep[JADES][]{Eisenstein2023_jades_overview}, Mirage or Miracle \citep[MoM][]{Naidu2026}, Red Unknowns: Bright Infrared Extragalactic Survey \citep[RUBIES][]{deGraaff2025_RUBIES}, Ultradeep NIRSpec and NIRCam Observations before the Epoch of Reionization \citep[UNCOVER][]{Price2024_uncover_spectra} and numerous others. A full list of all the spectroscopic surveys used in this work is provided in Table \ref{tab:spec-surveys} in the Appendix.

\subsection{Matching DJA spectroscopy and photometry}
\label{subsec:spec_phot_match}

The photometric counterparts of the \jwst/NIRSpec PRISM spectra on the DJA are found with a simple [RA, Dec] criterion.
The [RA, Dec] of the spectroscopic sources correspond to the positions in the original catalogue used to plan the observations and are retrieved from the MSA metadata. Photometric counterparts are found within an 0.5$^{\prime\prime}$ search radius from the spectroscopic source and the nearest match is stored. The median offset is 0.02$^{\prime\prime}$ and 95\% of the distribution have offsets less than 0.13$^{\prime\prime}$.

\subsection{Sample selection from the DJA}
\label{subsec:sample_selection}

We begin with a sample of all the publicly available and reduced NIRSpec PRISM v4.5 spectra available on the DJA. All spectra are visually inspected and assigned grades of 1, 2 or 3 depending on the quality of the spectrum, with 3 being the highest quality, with a visually approved spectroscopic redshift from an emission line or continuum feature \citep[see][for details on the grading scheme]{deGraaff2024b}. We select spectra assigned a grade of 3 that have a spectroscopic redshift of $z\geq2$. This {\tt grade=3} criterion removes contaminants such as stars and very low SNR sources. Further, we require that all spectra have at least one photometric counterpart (see previous subsection).
With these criteria, we identify a parent sample of $\sim 12,000$ spectra across all spectroscopic redshifts beyond 2.

As our sample is intended to isolate galaxy-dominated spectra, we next flag interlopers that can pass our selection criteria while not being straightforwardly interpretable as galaxies.  This includes Little Red Dots (LRD), which are compact objects with V-shaped rest-frame optical SEDs that were discovered in early \jwst grism data \citep{Labbe2023a, Matthee2023, Kocevski2023_pub} and often masquerade as high-redshift massive galaxies \citep{Labbe2023a}. We flag them in our sample based on the study by \cite{deGraaff2025_LRDoverview}, which conducted a systematic search for LRDs throughout the DJA on {\tt grade=3} NIRSpec/PRISM spectra, finding 147 LRDs at $2.0<z<9.3$. We additionally cross-match our catalogue with the LRD catalogues presented in \cite{Hviding2025} (spectroscopic LRD selection at $z>3.1$ in the RUBIES survey), \cite{Kokorev2024} (photometric LRD selection at $z \sim 4-9$) and \cite{Kocevski2025} (photometric LRD selection at $z \sim 2-11$), finding 35, 67 and 109 individual spectroscopic matches, respectively. Given the uncertain physical origin and complex rest-frame optical continuum shapes of LRDs, our modelling method struggles to reproduce their SEDs; we therefore flag these sources in our sample.

In order to identify broad-line active galactic nuclei (BLAGN), we flag sources based on the following studies. (i) \cite{Baccus2026} did a systematic search for BLAGN through $\sim80,000$ medium and high resolution grating spectra on the DJA, finding 227 BLAGN at $0.8<z<7.2$, (ii) \cite{Juodvbalis2026} searched for BL-AGN in the JADES survey from $1.5<z<9$, and (iii) broad Balmer line sources were identified in \cite{Hviding2025}. We cross-match these samples with our sample of PRISM spectra using an 0.2'' search radius and identify 110, 69 and 31 matches in our sample from \cite{Baccus2026}, \cite{Hviding2025}, and \cite{Juodvbalis2026}, respectively. Given that our modelling method does not include a AGN component, we flag these BLAGN and remove them in our galaxy-focused analyses in the following sections of this paper.

We identify strong gravitationally lensed spectra as those with photometric matches in the following Frontier Fields \citep{Lotz2017}: Abell 2744, Abell 370, MACS J0416.1-2403, MACS J0417.5-1154, MACS0647+7014, MACS J1149.5+2223 and MACSJ1423.8+2404 \citep[see][for an overview of the photometry]{Valentino2023}. In total, we identify 1317 sources in these fields. Given the need for a magnification maps to correctly interpret the stellar masses and SFRs of these sources, we flag them in our catalogue and do not include them in the results of this paper (these sources are however included in the data release for future use in context of the magnification maps for these fields).

\begin{figure*}[t!]
\includegraphics[width=\linewidth]{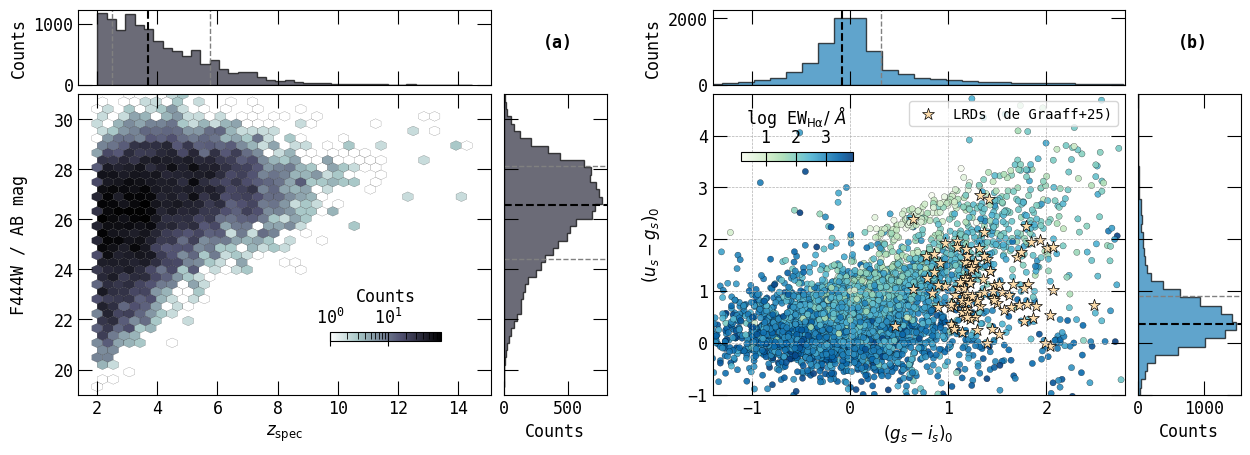}
\caption{The empirical landscape of our sample: {\sl (a)} AB magnitude in the F444W band vs. spectroscopic redshift for the full parent sample of 11,000 galaxies. The sample spans a redshift range of $2<z_{\rm spec}<14.4$ and a magnitude range of $20<{\rm mag_{F444W}}<32$ in the F444W filter. The black dashed line on the histograms indicates the median value of the distribution and the grey dashed lines indicate the $16^{\rm th}$ and $84^{\rm th}$ percentile values. Most of our sample lies below $z_{\rm spec}=6$ and spans an F444W magnitude range of $24-28$ mag.
{\sl (b)} a \ugis{} rest-frame colour-colour diagram of our sample, coloured by the logarithmic $H\alpha$ EW. LRDs from \cite{deGraaff2025_LRDoverview} are denoted by stars in each plot. In light green, we see a sequence of galaxies with low $H\alpha$ EW.}
\label{fig:sample_selection_1}
\end{figure*}

Figure \ref{fig:sample_colour_space_0} shows the colour-magnitude-redshift space spanned by the spectroscopic sample used in this work, compared with the photometric sample. The sources in the photometric sample shown here are from the GOODS-North and GOODS-South fields (specifically catalogues {\tt gdn-grizli-v7.4} and {\tt gds-grizli-v7.2} on the DJA; see Table \ref{tab:phot-cats} for details). The sources shown here are chosen to have photometric redshifts of $z_{\rm phot}>2$ for both the spectroscopic and photometric samples. The F444W magnitude range for both samples is restricted to $18<F444W<30.5$ (the faint-end limit corresponds to the average $1\sigma$ depth of the photometric catalogues).

The first panel of Figure \ref{fig:sample_colour_space_0} shows that the spectroscopic sample spans a similar $z_{\rm phot}$ range as the photometric sample, and overall traces brighter F444W magnitudes. The central and final panels shows that both samples span a similar F150W-F444W colour space as a function of redshift as well as a similar colour-magnitude space.

An overview of our full parent sample is shown in Figure \ref{fig:sample_selection_1}. Our sample of $\sim 12,000$ spectra span a redshift range of $2<z<14.4$ and a magnitude range of $20<{\rm mag_{F444W}}<32$ in the NIRCam F444W filter (centred at 4.4 ${\rm \mu m}$). A majority of our sample lies at $z<6$ and occupies a magnitude range of $24-28$ mag.

The right-hand panel in Figure \ref{fig:sample_selection_1} shows our sample scattered on the rest-frame (spectroscopically derived) \ugis{} colour-colour diagram, shown to efficiently separate quiescent from star-forming galaxies \citep{Danso2023}. The data are coloured by H$\alpha$ EW (from the DJA). Galaxies with the highest EW correspond to the bluest \ug{} and \gi{} colours, presumably star-forming galaxies with young stellar populations. Galaxies with the lowest EW carve out a tight sequence in this colour space, amongst the reddest \ug{} colours of this sample; these are presumably galaxies which host old stellar populations and have little-to-no current star-formation. Our subsequent modelling in this paper will seek to characterise the stellar population properties of these galaxies.

\subsubsection{Completeness of the DJA spectroscopic database}
\label{subsec:sample_incompleteness}

The DJA is a public repository containing data from hundreds of individually designed programmes targeting galaxies with NIRSpec/PRISM. Programmes such as JADES \citep{Eisenstein2023_jades_overview}, GTO-WIDE \citep{Maseda2024}, MoM \citep{Naidu2026} and CAPERS \citep{Dickinson2024} have focused on high redshift selections of UV bright targets with an emphasis on pushing the redshift frontier as far back as possible; other surveys such as RUBIES \citep{deGraaff2025_RUBIES} have aimed at a selection in colour spaces that represent underlying photometric colour-magnitude-redshift distributions with a well-defined and reproducible selection function. However, even spectroscopic surveys that aim for completeness are inherently limited by what is observable with the MSA in terms of minimum source brightness and shutter placement conflicts.

This study focusses on SED modelling all high-quality and visually-vetted spectroscopically observed sources (with PRISM) at $z_{\rm spec}>2$.
Since our study draws from all public surveys individually designed for heterogeneous scientific aims, we acknowledge the potential biases inherent to this sample.
We caution that future sample studies that address relative abundances and redshift evolution must account for the heterogeneity of this selection.

\subsection{Final sample}
\label{subsec:final_sample}

Our full parent sample comprises of $\sim 12,000$ galaxies from redshifts of $2<z_{\rm spec}<14.4$. The modelling described in the following section is performed on all spectra in this parent sample, and the results are released for all modelled spectra. For the results presented from Section \ref{sec:results} onwards, we leave out the LRDs identified in \cite{deGraaff2025_LRDoverview, Hviding2025, Kokorev2024, Kocevski2025}, and BL-AGN identified in \cite{Baccus2026, Hviding2025, Juodvbalis2026}. We also leave out all spectra lying in lensing fields, as previously described. Further, we exclude spectra that have fewer than 450 valid pixels out of a total of 469 pixels from the latest v4.5 DJA spectral reduction (1614 spectra), so as to get largely uniform spectral coverage from 0.5 to 5.5 ${\rm \mu m}$. Finally, we also exclude spectra that have fewer that 4 photometric bands (74 spectra). The final sample discussed from Section \ref{sec:results} onwards comprises $\sim 9,000$ spectra.

\begin{table*}[!htb]
{\renewcommand{\arraystretch}{1.5} 
\caption{The priors set on the parameters used in our Bagpipes SED modelling setup.}
\label{tab:pipes-priors}

\resizebox{\textwidth}{!}{%
\begin{tabular}{l l l l l l l}
\hline
\hline
{\bf Galaxy component} & {\bf Parameter} & {\bf Units} & {\bf Prior distribution} & {\bf Range} & {\bf Hyperparameters} & {\bf References}\\
\hline
Star-formation history & $\rm{\Delta SFR = \log (SFR_{n} / SFR_{n+1})}$ & & Student's t & (-10.0,10.0) & $\sigma=0.3,\ \nu=2$ & \cite{Leja2019} \\
(continuity)  & Formed stellar mass, $\rm{M_{formed}}$ & $\rm{M_\odot}$ & Logarithmic & $(10^6, 10^{14})$ & & \\
  & Stellar metallicity, $\rm{Z_\star}$ & $\rm{Z_\odot}$ & Uniform & (0.01, 2.5) & & \\
\hline
Dust  & $V$-band attenuation, \av & mag & Uniform & (0.0, 6.0)  & & \cite{Salim2018_dust_law} \\
  & Deviation from Calzetti slope, $\delta$ & & Gaussian & (-1.0, 0.3) & $\mu=0,\ \sigma=0.1$ & \cite{Carnall2024} \\
  & 2175\AA\ UV bump strength, $B$ & & Uniform & (0, 5) & & \\
\hline
Nebular  & Ionisation parameter, $\log U$ & & Gaussian & ($-4,\ -1$) & $\mu=-2.25,\ \sigma=0.25$ & \\
\hline
Calibration  & $0^{\rm th}$ order, $\mathcal{C}_0$ & & Gaussian & $(C_{0,\mu} \pm 5\times C_{0,\sigma})$ & $\mu=C_{0,\mu},\ \sigma=C_{0,\sigma}$  & \\
  & $1^{\rm st}$ order, $\mathcal{C}_1$ & & Gaussian & $(C_{1,\mu} \pm 5\times C_{1,\sigma})$ & 
  $\mu=C_{1,\mu},\ \sigma=C_{1,\sigma}$  & \\
  & $2^{\rm nd}$ order, $\mathcal{C}_2$ & & Gaussian & $(C_{2,\mu} \pm 5\times C_{2,\sigma})$ & $\mu=C_{2,\mu},\ \sigma=C_{2,\sigma}$  & \\
\hline
Redshift & $z_{\rm spec}$ & & Gaussian & ($z_{\rm spec} \pm 0.005\times(1+z_{\rm spec})$) & $\mu=z_{\rm spec}$ & \\
  & & & & & $\sigma=0.001\times(1+z_{\rm spec})$ & \\
\hline
Noise  & White noise scaling & & Logarithmic & (0.01, 10.0) & & \cite{Carnall2023} \\
\hline
Fixed parameters  & Velocity dispersion & ${\rm km/s}$ & & & 100.0 & \\
& Birth cloud age, $t_{\rm BC}$ & Myr & & & 25.0 & \\
  & Birth cloud attenuation ratio, $\epsilon$ & & & & 2.0 & \cite{Witstok2023_2175_at_z7}\\
\hline
\hline
\end{tabular} 
}
}
\end{table*}

\section{SED modelling with \bagpipes{}}
\label{sec:SED_modelling}

We perform joint spectrophotometric SED modelling with the Bayesian Analysis of Galaxies for Physical Inference and Parameter
EStimation \citep[\bagpipes{},][]{Carnall2018, Carnall2019} software. \bagpipes{} is a stellar population synthesis (SPS) code which uses Bayesian inference to build and fit galaxy SEDs to photometric and spectroscopic data. It utilises the {\tt nautilus} sampling algorithm for Bayesian posterior estimation \citep{Lange2023_nautilus}.
We use \bagpipes{} to generate model SEDs of galaxies using SPS models built on the BPASS v2.2.1 stellar spectral library, with the ``135\_300'' broken power-law initial mass function (IMF) as described in \cite{Stanway2018}: a power-law index $\alpha_1=-1.30$ in the mass range $M_\star \in (0.1,0.5)\ M_\odot$ and $\alpha_2=-2.35$ in the range $M_\star \in (0.5,300)\ M_\odot$ \citep[see an implementation of these grids in][]{Giovinazzo2026}. The code incorporates user-specified SFHs, nebular emission modelling from the photoionisation code \cloudy{} \citep{Ferland2017}, user-specified dust attenuation laws, and an IGM attenuation model from \cite{Inoue2014}.

We describe our SED modelling set-up below. Notably, we introduce a flexible emission line marginalisation technique, described in Section \ref{subsec:line_marginalisation}; see Section \ref{subsec:djapipes-eval} for a discussion on the impact of using this method on the fitting results. A summary of the parameters and priors used in the modelling can be found in Table \ref{tab:pipes-priors}.

\subsection{Star-formation history}
\label{subsec:sfh_model}

We use the non-parametric continuity SFH prior as described in \cite{Leja2019}, which allows for flexible SFRs within defined time-steps. The continuity prior places constraints on the logarithmic ratio of the SFRs between adjacent time bins (${\rm \Delta SFR = \log (SFR_{n} / SFR_{n+1})}$), informed by the student's t-distribution, where we centre the distribution at 0 with a scale factor $\sigma = 0.3$ (which controls the width of the distribution) and $\nu = 2$ \citep[see][]{Leja2019, Tacchella2022}. We create a fixed grid of time steps in lookback time, with the first few time bin edges fixed at [0, 3, 10, 25, 50, 100] Myr. Thereafter, bin edges are defined in 10 equally spaced intervals in logarithmic lookback time ranging from 100 Myr to the age of the universe at ${z=0}$ minus the age at ${z=30}$ (which corresponds to $\sim100$ Myr after the Big Bang). On this fixed grid, the last time step is flexibly determined by the age of the universe at the spectroscopic redshift of the source. The definition of time-steps for any source is thus the same in lookback time (except for the last time step), while the number of time-steps depends on the redshift of the source, ${\rm z_{spec}}$. A maximum of 14 time bins are allowed $z=0$.

We allow the stellar mass formed (as opposed to that retained) to vary within ${\rm M_\star / M_\odot \in (10^{6}, 10^{14})}$ with a flat logarithmic prior and the metallicity to vary within ${\rm Z_\star \in (0.01, 2.5)\ Z_\odot}$ with a uniform prior.

\subsection{Nebular model}
\label{subsec:neb_model}

We use \cloudy{} to generate nebular line and continuum models using the BPASS stellar models as the input ionising source, excluding internal ISM dust modelling within \cloudy{} and extending the number of emission lines in the models to include some far UV lines such as C\,{\sc iii} and C\,{\sc iv} \citep[as implemented in][]{Giovinazzo2026}. These changes have now been propagated to the latest version of \bagpipes{}.

The nebular ionisation parameter is varied within a range of ${\log U \in (-4, -1)}$ with a Gaussian prior centred at -2.25 with a standard deviation of 0.25; while local H\,{\sc ii} regions typically have ionisation parameter values in the range ${ \log U \in (-3.5, -2.5)}$ \citep{Dopita2000, Kewley2019}, studies show that for strongly star-forming and bursty galaxy environments, $\log U$ can be as extreme as -1.5 \citep[e.g.,][]{Reddy2023}.

We additionally refer the reader to Section \ref{subsec:line_marginalisation} where we describe a line marginalisation method to account for under-/over-estimated line fluxes from the \cloudy{} grids.

\begin{figure*}[htb!]
\includegraphics[width=0.9\linewidth]{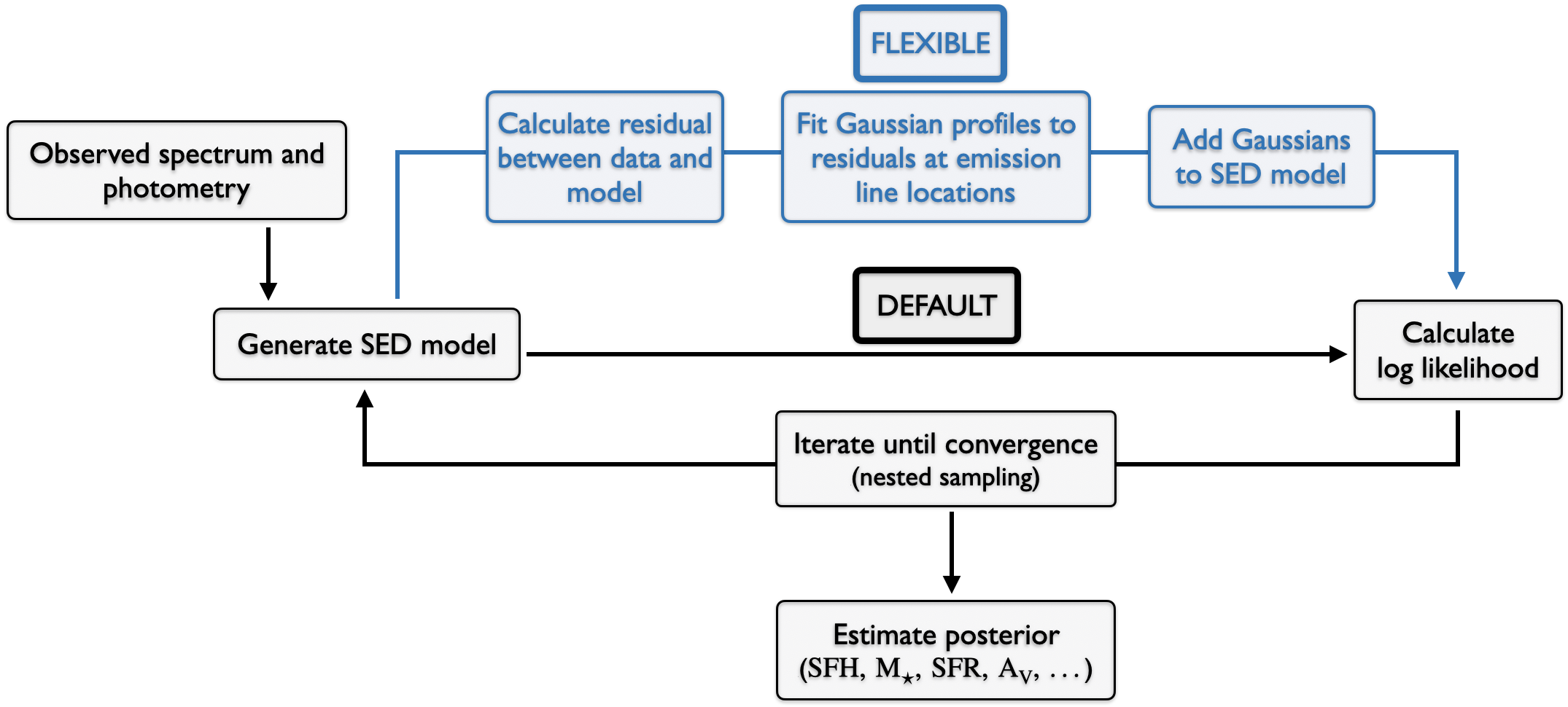}
\caption{A schematic displaying our custom fitting method and how it differs from default fitting within \bagpipes{}. The blue arrow and boxes describe the additional steps taken within the flexible fitting procedure to marginalise over the emission line profiles. At every instance that an SED forward model is generated during the exploration phase of the fitting, we fit Gaussian models to the residual between the model and the data to capture any additional emission that the SED model may not have reproduced. In doing so, our method allows the stellar continuum of the spectrum to drive the fit while being agnostic to the origin or nature of the emission lines.}
\label{fig:bagpipes-mod-flow-chart}
\end{figure*}

\subsection{Dust model}
\label{subsec:dust_model}

We use a \cite{Salim2018_dust_law} dust attenuation law, parametrised by ${\rm \delta}$, the power-law deviation slope from the \cite{Calzetti2000} attenuation curve, and $B$, the strength of the $2175 \text{\AA}$ bump \citep[an extinction feature due to carbonaceous grains;][]{Stecher1965_UV_bump}. The deviation $\delta$ is varied in the range ${\rm \delta \in (-1.0, 0.3)}$ with a Gaussian prior centred on $\mu=0.0$ with a standard deviation $\sigma=0.1$ \citep[see][]{Carnall2024}. The lower limit range for $\delta$ is chosen to allow for extremely steep slopes, as found for some galaxies in \cite{Salim2018_dust_law}. The prior is centred at $\delta=0.0$, which corresponds to a \cite{Calzetti2000} attenuation curve; this is chosen such that the dust law defaults to this when the data is not constraining enough. $B$ is varied in the range ${B \in (0, 5)}$, also spanning the range of values found in \cite{Salim2018_dust_law}.
The $V$-band (${\rm \sim 5500} \text{\AA}$) dust attenuation is varied in the range ${\rm A_V \in (0, 6)\ mag}$ with a uniform prior. We include a multiplicative factor on the \av{}, $\epsilon=2$ \citep[as in][]{Carnall2019, Witstok2023_2175_at_z7} that is applied to emission from stars within stellar birth clouds, with the maximum age of birth clouds fixed to 25 Myr \citep{Carnall2019,Salim2020_dust_review}.

\subsection{Wavelength-dependent resolution of PRISM}
\label{subsec:wav_resolution}

Given the wavelength-dependent resolution of the \jwst/NIRSpec PRISM instrument, ranging from a resolving power of $R\approx30$ to $R\approx330$ from ${\rm 0.6 - 5.3\ \mu m}$ \citep{Jakobsen2022}, the model spectra need to be resampled onto an appropriate wavelength grid. Thus, to convolve the model spectra, we use the PRISM resolution curve specified in the \jwst User Documentation (JDox)\footnote{\protect \url{https://jwst-docs.stsci.edu/}} and include a multiplicative factor of 1.3, which accounts for the higher resolution of compact sources \citep[which are more common at higher redshifts; see][]{deGraaff2024_kinematics}. Given the low resolution of PRISM spectroscopy, our data do not have the constraining power to determine stellar or ionised gas velocity dispersions below $\approx {\rm 750\ km/s}$\footnote{\protect\url{https://jwst-docs.stsci.edu/jwst-calibration-status/nirspec-calibration-status/nirspec-mos-calibration-status\#gsc.tab=0}}. We therefore fix the velocity dispersion to 100 km/s, a typical value for high redshift galaxies \citep{deGraaff2024_kinematics}.

\subsection{Photometric calibration}
\label{subsec:phot_calibration}

We perform joint spectrophotometric SED modelling, with the spectra calibrated to the photometry with a second-order Chebyshev polynomial, which is built into \bagpipes{} within the Bayesian framework. In order to set informed priors on the three polynomial coefficients of the calibration curve, we perform an initial least-squares fit of a second-order Chebyshev polynomial to the ratio between the observed photometry and the synthetic photometry. The synthetic photometry is calculated by integrating the observed spectrum through the appropriate filter curves. The coefficients resulting from this least-squares fit and their associated uncertainties are used to define Gaussian priors on the coefficients. Table \ref{tab:pipes-priors} shows the exact prescriptions.

\subsection{Other parameters}
\label{subsec:nuisance_params}

We adopt a tight Gaussian redshift prior centred around $z{\rm _{spec}}$ with a standard deviation of $ 0.001\times(1+z_{\rm spec})$, where $z{\rm _{spec}}$ is calculated with \msaexp{} and available along with the reduced spectra on the DJA.

To account for potential over- or under-estimation of the spectral noise, we place a logarithmic prior in the range (0.1, 10) on the constant factor by which the spectral noise is scaled \citep[as in][]{Carnall2024}.

\begin{figure*}[t!]
\includegraphics[width=\linewidth]{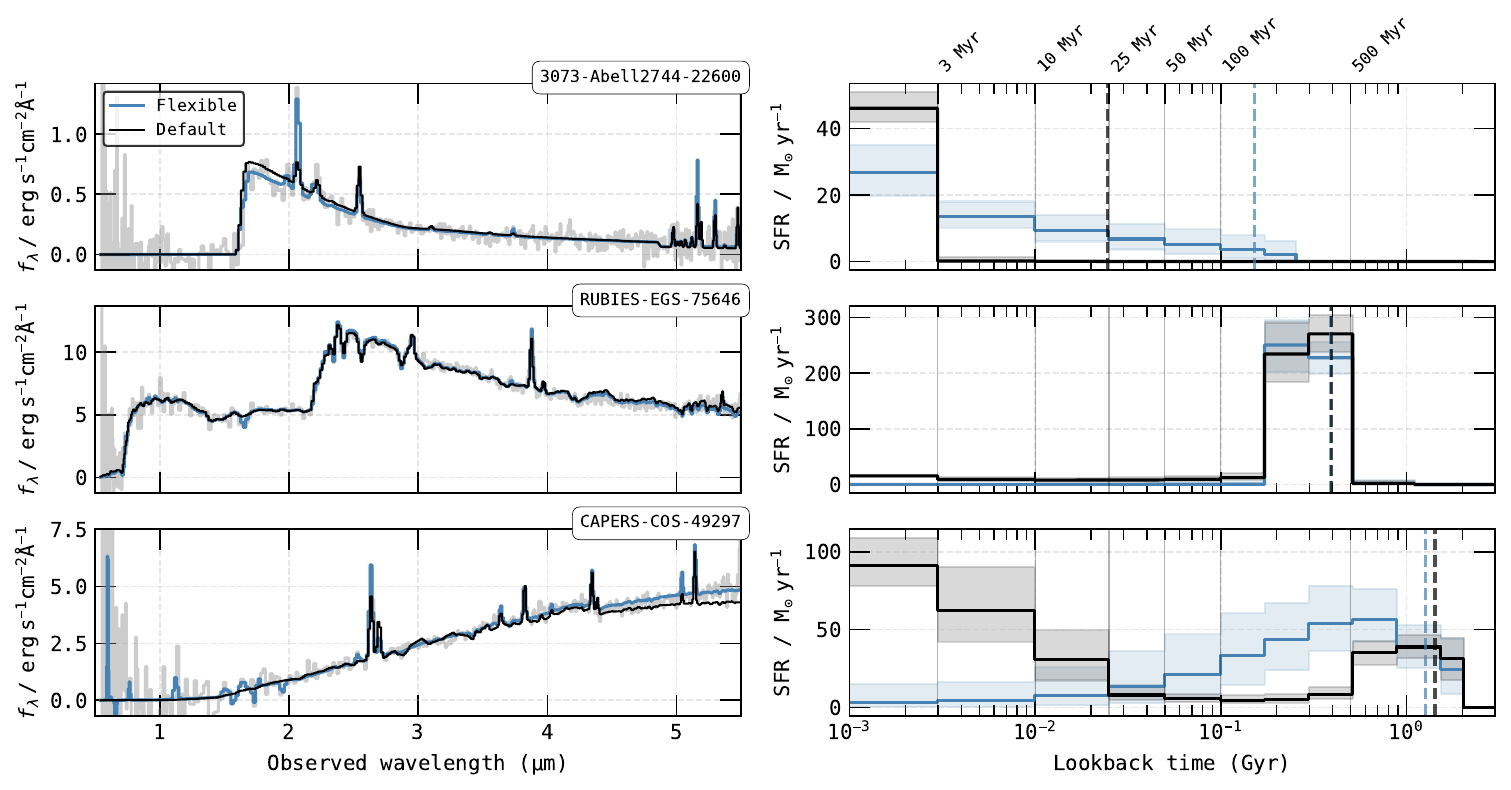}
\caption{Example of flexible (blue) and default (black) \bagpipes{} SED fits and SFHs of a star-forming galaxy at $z=12.3$, a quiescent galaxy at $z = 4.9$ and a dusty galaxy at $z=3.0$. For the star-forming galaxy, emission lines are better captured by the flexible line fitting method, which also prefers a gradually rising SFH as opposed to the rapid SFH preferred by the default fitting method. For the quiescent galaxy, while both SED fits are comparable, the flexible method prefers a completely quiescent SFH solution while the default SFH finds some residual star-formation in the most recent time bins, arising from mis-identifying an [N\,{\sc ii}] line as H$\alpha$ due to the PRISM resolution. In the case of the dusty galaxy, both recent SFHs look dramatically different; we see that the flexible line fitting captures the emission lines better as well as the NIR stellar continuum.}
\label{fig:technical_seds_flex_vs_def}
\end{figure*}

\subsection{Marginalisation over emission lines}
\label{subsec:line_marginalisation}

In initial tests on our sample, the SED models with the set-up described above were unable to reproduce the observed EW for certain strongly star-forming galaxies. This often resulted in (i) underestimating the observed EW of emission lines such as ${\rm H\alpha}$ or [O\,{\sc iii}]${\lambda\lambda4959,5008}$, while also (ii) overestimating the continuum around these lines, simultaneously (iii) not fitting certain emission lines such as [O\,{\sc ii}]${\lambda\lambda 3727,3730}$, for a typical outcome where (iv) the SFH drives the fit to the youngest possible ages (pushing the entire SFH into the youngest time bin) to reproduce the high EWs.

To counter this effect, we develop a custom version of \bagpipes{}\footnote{\protect\url{https://github.com/RashmiGot/bagpipes}}. This marginalizes over emission lines by fitting flexible Gaussians to the \bagpipes{} models at each instance where a model is generated in the fitting framework. A similar nebular marginalisation functionality is implemented in {\sc Prospector} \citep[see][for an implementation]{Zhang2025}.

This is shown in the schematic in Figure \ref{fig:bagpipes-mod-flow-chart}: at each forward model call, we fit Gaussian profiles using the least-square method at known emission line locations to the residual between the model (down-calibrated to match the normalisation of the spectroscopic data) and the data and add these line profiles to the forward model before estimating the likelihood. The \cloudy{} modelling and flexible Gaussian modelling of the emission lines are therefore mutually exclusive. A list of all the emission lines used in the flexible line fitting model is provided in Table \ref{tab:line-wavelengths} in the Appendix.
In the case of doublets such as [O\,{\sc iii}]${\lambda\lambda4959,5008}$ or nearby/blended lines such as H${\alpha}$+[N\,{\sc ii}], fixed line ratios are used as defined in \msaexp{}.

This method makes our fits agnostic to the presence and strength of emission lines such that the modelling prioritises fitting the stellar continuum over the line regions of the spectrum. It also allows us to recover the full line fluxes of emission lines by combining the fluxes predicted from the \bagpipes{} models to those found by our least-squares fitting method. The efficacy of this custom fitting method is evaluated below.

\begin{figure*}[t!]
\includegraphics[width=0.8\linewidth]{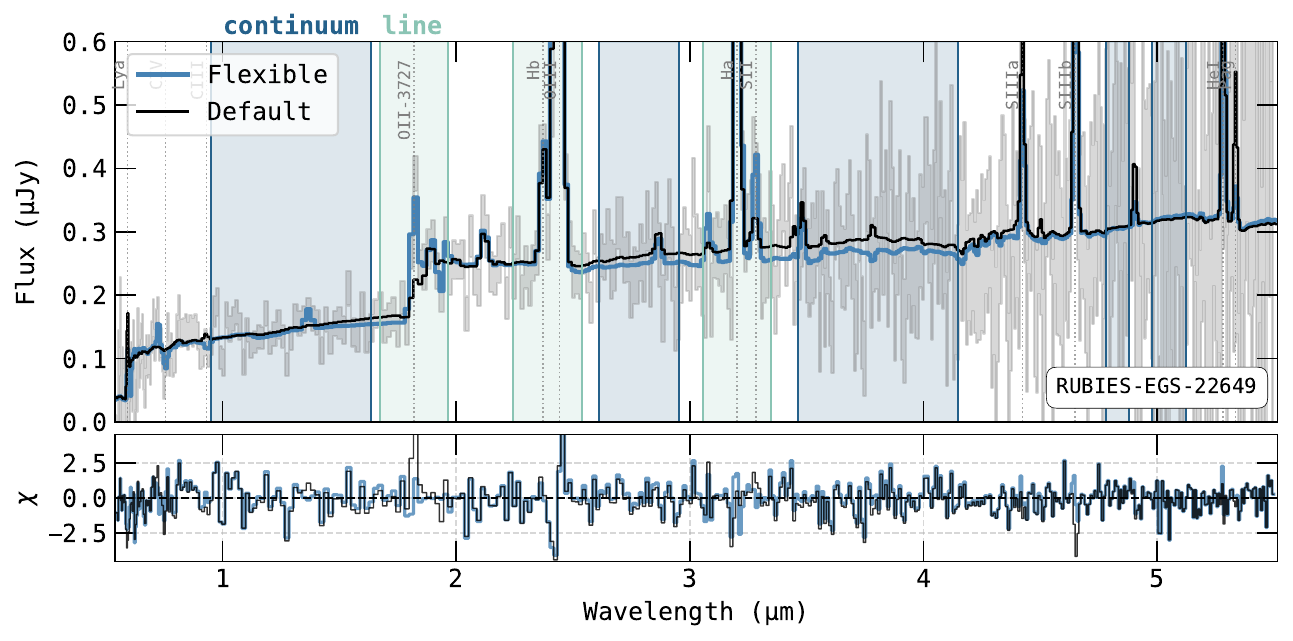}
\includegraphics[width=0.8\linewidth]{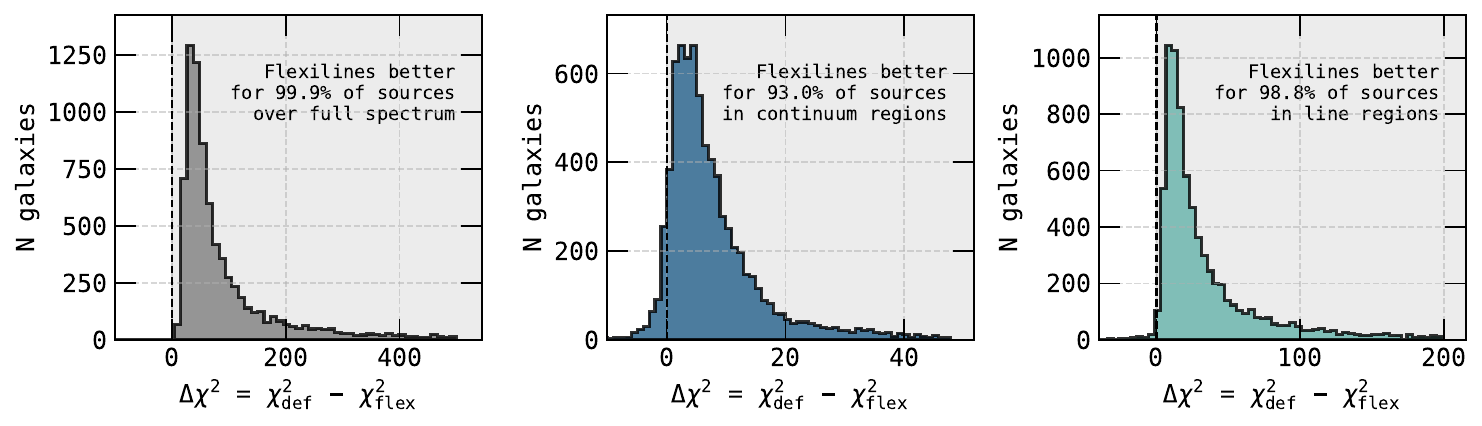}
\caption{{\sl Top:} example of flexible and default \bagpipes{} SED fits to {\tt RUBIES-EGS-22649} \citep{deGraaff2024b}, with the highlighted sections indicating the continuum regions (blue) and line regions (green) within which the $\Delta \chi^2$ are compared. In addition to affecting the line modelling (see e.g. region around [O\,{\sc iii}]${\rm \lambda 3727}$), the flexible line modelling affects the stellar continuum as well. {\sl Bottom:} the $\Delta \chi^2$ distributions calculated over the full spectrum (grey), the continuum regions (blue) and the line regions (green), where $\Delta \chi^2 > 0$ implies that $\chi^2_{\text{default}} > \chi^2_{\text{flexible}}$. By construction, our flexible method performs better in the line regions for 98\% of sources, but importantly also improves the modelling in the stellar continuum regions for 93\% of all sources in our sample.}
\label{fig:sedmodelling_chi2}
\end{figure*}

\subsection{Evaluating the fitting method}
\label{subsec:djapipes-eval}

In order to evaluate the performance of our custom fitting method, we fit all galaxies in our sample with the ``default'' method (i.e. \bagpipes{} modelling excluding the flexible line modelling) and the ``flexible'' method (which includes the flexible line modelling functionality). We compare goodness of fits and posterior physical properties of these sources between the flexible fitting and the default \bagpipes{} fitting. Keeping all else the same (priors on e.g., the SFH, dust, nebular emission and calibration), the only difference between the fitting methods is turning on the additional emission line fitting described in Section \ref{subsec:line_marginalisation} and Figure \ref{fig:bagpipes-mod-flow-chart}.

Figure \ref{fig:technical_seds_flex_vs_def} shows examples of three galaxies fit with both methods: a $z=12.3$ star-forming galaxy (\citep[\texttt{\seqsplit{3073-Abell2744-22600}};][]{Castellano2024}, a $z=4.9$ quiescent galaxy \citep[\texttt{\seqsplit{RUBIES-EGS-75646}};][]{deGraaff2024b} and a $z=3.0$ dusty star-forming galaxy \citep[\texttt{\seqsplit{CAPERS-COS-49297}};][]{Dickinson2024}. For the star-forming galaxy, the emission lines are better captured by the flexible line-fitting method, and the best-fit SFH prefers a gradually rising solution, as opposed to the default method where all the star-formation is piled up into the most recent 3 Myr time bin. There are also notable differences in the stellar continuum, especially in the far-UV. For the quiescent galaxy, the flexible method prefers a SFH with zero star formation in the last few hundred Myr, while the default method contains residual star formation in the most recent time bins. For the dusty galaxy, the NIR continuum shape is better captured by the flexible method, and the best-fit SFH prefers a gradually falling solution while the default fit prefers a steeply rising recent SFH.

While Figure \ref{fig:technical_seds_flex_vs_def} focuses on some individual examples, Figure \ref{fig:sedmodelling_chi2} shows a comparison between the $\chi^2$ values of the two fitting methods for our whole sample. We identify regions in a spectrum dominated by the stellar continuum and those by certain emission lines (e.g. H${\alpha}$+[N\,{\sc ii}], ${\rm H\beta}$+[O\,{\sc iii}], [O\,{\sc ii}]${\rm \lambda 3727}$). We then compare the $\chi^2$ values across (a) the full fitted spectrum, (b) the fitted spectrum in the continuum regions and (c) the fitted spectrum in the line regions using the quantity $\Delta \chi^2 = \chi^2_{\text{def}} - \chi^2_{\text{flex}}$, where if $\Delta \chi^2 > 0$ then the flexible line fitting method yields a smaller $\chi^2$ than the default method. In each case, the fits with the flexible line modelling outperform the default fitting methods for an overwhelming majority of our sample. In the overall spectrum, the modified method yields better fits for $>99\%$ of sources, compared with $93\%$ ($98.8\%$) of sources in the continuum (line) region alone.

\begin{figure*}[htb!]
\includegraphics[width=\linewidth]{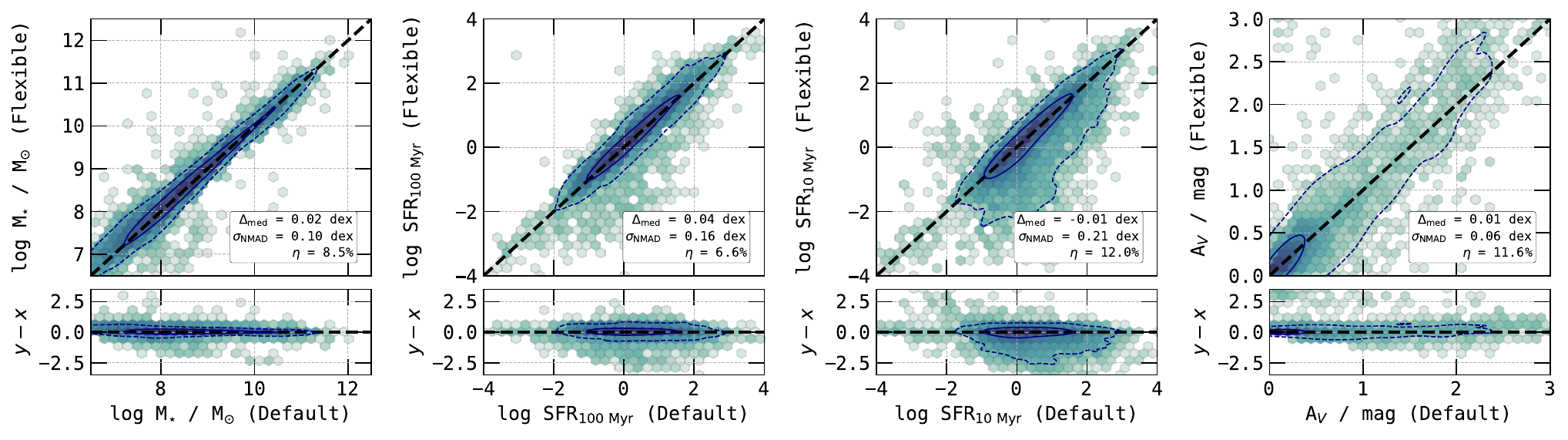}
\caption{Comparison of key physical properties between the flexible ($y$ axis) and default ($x$ axis) fitting methods. From left to right, the panels show comparisons of ${\rm M_\star}$, ${\rm SFR_{100\ Myr}}$, ${\rm SFR_{10\ Myr}}$ and \av{}. The lower panels show the differences between the two fitting methods. Contours represent the $1\sigma$ and $2\sigma$ levels. While stellar masses and star formation rates on 100 Myr timescales are in tight agreement overall, we see that star formation rates on 10 Myr timescales tend to scatter below the 1:1 relation (see 2$\sigma$ contour). Also quoted within the plots are the median offset ($\Delta_\text{med}$), the Normalized Median Absolute Deviation (NMAD; $\sigma_\text{NMAD}$) and the outlier fraction ($\eta$), defined the percentage of sources lying more than $3\times\sigma_\text{NMAD}$ away from the median offset.}
\label{fig:flex_vs_def_physical_props_v0}
\end{figure*}

Figure \ref{fig:flex_vs_def_physical_props_v0} shows a comparison of some posterior physical properties between standard \bagpipes{} fitting and our custom version. The stellar masses (${\rm \sigma_\text{NMAD} = 0.1\ dex}$) and star-formation rates averaged over 100-Myr timescales (${\rm \sigma_\text{NMAD} = 0.16\ dex}$) are in good agreement, while for the star-formation rates on 10-Myr timescales (${\rm \sigma_\text{NMAD} = 0.21\ dex}$), some outliers tend to scatter below the 1:1 relation, implying that for some sources the flexible line fitting yields lower SFRs than the default fitting. This is to be expected as the flexible line SED fits (and therefore the SFHs) are agnostic to the presence of emission lines while for the default fitting method the fits and SFHs are directly tied to the strength of the emission lines. The V-band dust attenuation magnitudes are overall in good agreement but also show broad scatter in the $2\sigma$ contour; since the dust attenuation is also informed by the presence of emission lines, this scatter is not altogether surprising.

\begin{figure*}[t!]
\includegraphics[width=0.85\linewidth]{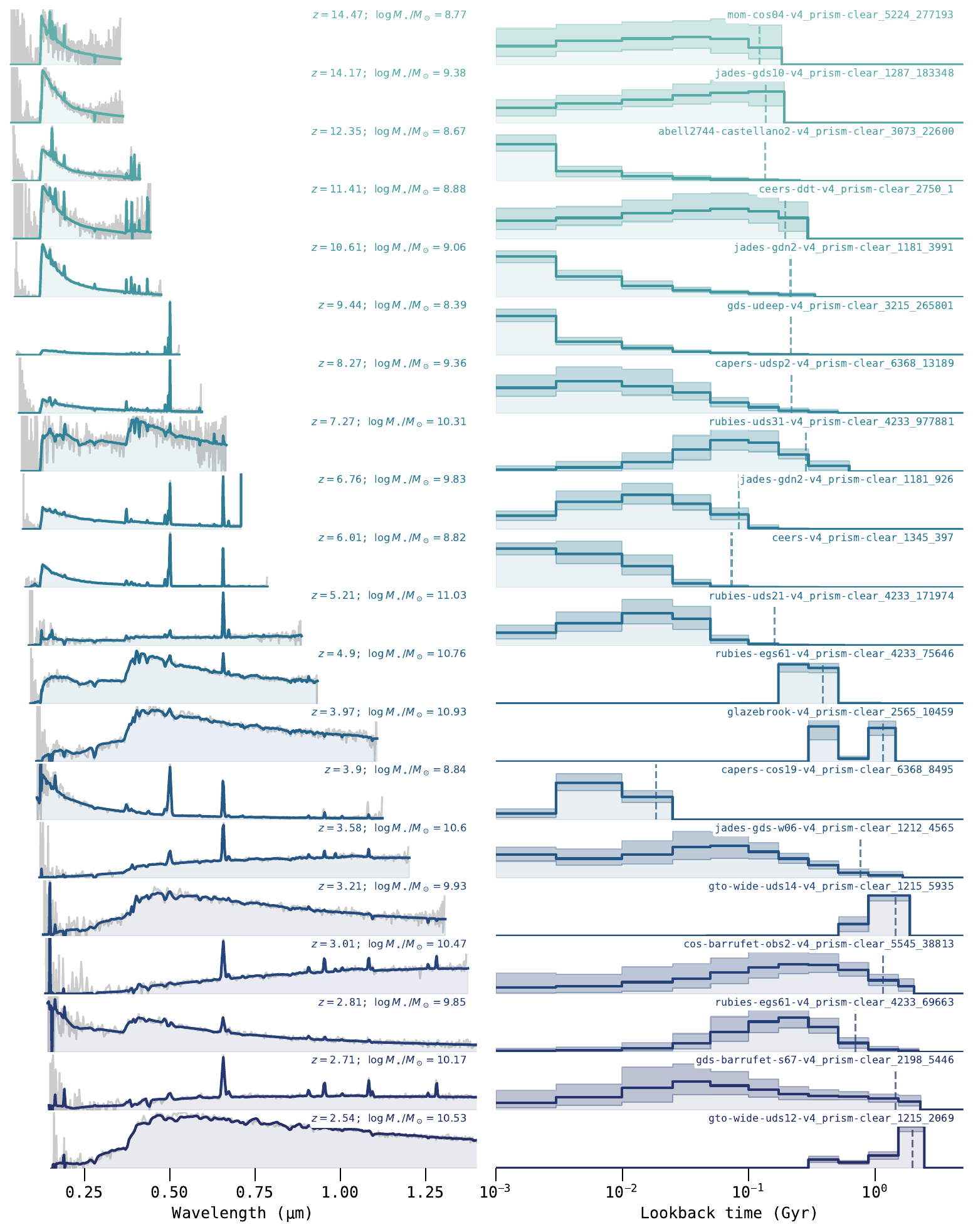}
\caption{Examples of SEDs (left) and SFHs (right) from our sample in the redshift range $2<z<14.4$, drawn from the following spectroscopic surveys: MoM \citep{Naidu2026}, JADES \citep{Eisenstein2025}, GO-3073 \citep{Castellano2024}, CEERS \citep{Finkelstein2023, ArrabalHaro2023_Nature}, CAPERS \citep{Dickinson2024}, RUBIES \citep{deGraaff2025_RUBIES}, GO-2565 \citep{Nanayakkara2024}, GTO-WIDE \citep{Maseda2024}, GO-5545 \citep{Barrufet2024_jwstprop} and GO-2198 \citep{Barrufet2024}. We see a significant diversity in the types of SEDs and SFHs in this sample, ranging from high redshift star-forming galaxies dominated by strong UV and optical emission lines to lower redshift galaxies dominated by old stellar populations and affected by dust extinction. A list of the stellar population properties of these galaxies can be found in Table \ref{tab:phys_props}.}
\label{fig:sedmodelling_specfits}
\end{figure*}

To validate the physical properties inferred from our spectrophotometric fits, we run a series of additional tests, with the full details presented in the appendices. In Appendix \ref{appendix:stellar_mass_phot_vs_spec}, we discuss how our stellar mass estimates compare with photometry-only fits using the SED fitting code \eazy{}. We find an overall scatter of ${\rm \sigma_\text{NMAD} = 0.32\ dex}$, with \eazy{} finding higher masses overall from photometry-only fits compared to our \bagpipes{} spectrophotometric fits, with the distribution having an overall offset of 0.23 dex.
We also test the reliability of our derived line properties in Appendix \ref{appendix:ha_flux_eqw}, comparing the ${\rm H\alpha}$+[N\,{\sc ii}] line fluxes and EW from the default and flexible line fitting methods to the fluxes and EW derived from \msaexp{} on the DJA. Both methods yield line fluxes that are in overall agreement with those found from \msaexp{}; the H$\alpha$+[N\,{\sc ii}] line fluxes derived from the custom fitting method (${\rm \sigma_\Delta = 0.21\ dex}$) are in closer agreement with \msaexp{} than those from the default \bagpipes{} method (${\rm \sigma_\Delta = 0.34\ dex}$). For the H$\alpha$+[N\,{\sc ii}] EW, both fitted methods broadly agree with \msaexp{} values but show significant scatter, where the flexible line method has a scatter of ${\rm \sigma_\Delta = 0.39\ dex}$ and the default method a scatter of ${\rm \sigma_\Delta = 0.46\ dex}$.

\begin{figure*}[htb!]
\includegraphics[width=0.9\linewidth]{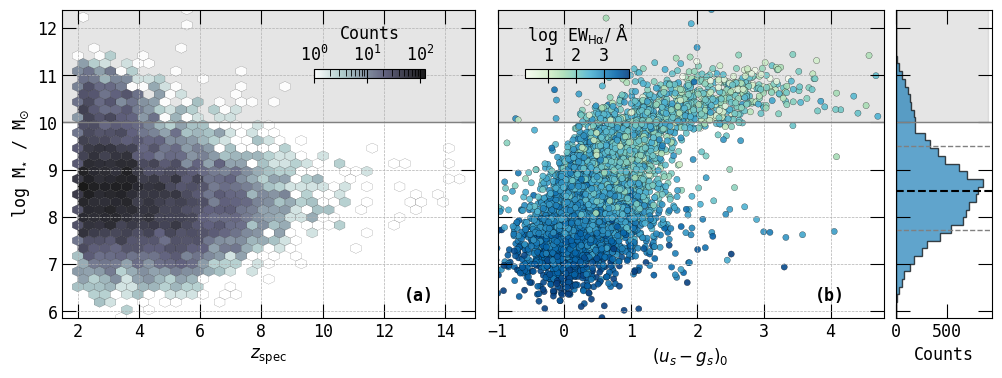}
\caption{Left: stellar mass vs. spectroscopic redshift for our SED-modelled sample. The sources span a mass range of $6 < \log M_\star/M_\odot < 12$ at redshifts of $z>2$, with a median stellar mass of $\log M_\star / M_\odot = 8$. Toward higher redshift, the upper limit of the most massive galaxies falls, which reflects the rarity of the most massive sources at high redshifts (spectroscopically observed). Right: stellar mass vs. \ug{} colour, coloured by H$\alpha$+[N\,{\sc ii}] EW. As stellar mass increases, galaxies transition from bluer \ug{} and \gi{} colours to redder colours.
}
\label{fig:results-zspec-vs-mass}
\end{figure*}

Overall, we find that the flexible line modelling implemented within \bagpipes{} yields better fitting results than the default fitting method (Figure \ref{fig:sedmodelling_chi2}) in terms of goodness-of-fit. While the flexible line modelling is, by construction, expected to yield better $\chi^2$ values both overall and in the line regions, we see that crucially this method also improves our modelling of the stellar continuum as the fits are no longer necessarily required to fit the emission lines. This highlights the fact that there is room of improvement in the default \cloudy{} grid approach, and points towards requiring more efficient nebular emission modelling in the future \citep[see][for a novel approach implemented in {\sc Prospector}]{Li2025}. This flexible modelling method has a direct impact on the inferred SFHs of sources (as shown in Figure \ref{fig:technical_seds_flex_vs_def}) and therefore on the derived physical properties of galaxies in this sample.

We acknowledge that this flexible modelling method could be interpreted as ``throwing away'' valuable information provided by emission lines, effectively similar to masking out these lines. Indeed, sources with strong emission lines that might be expected to have rising recent SFHs (for example, the dusty galaxy shown in Figure \ref{fig:technical_seds_flex_vs_def}) are instead fit with falling recent SFHs and overall older stellar populations with the flexible line modelling method. Quantities which are known to be traced by emission lines, such as ${\rm SFR_{10}}$, will also not be constrained by the emission lines themselves, but by the stellar continuum instead. However, since the default method of modelling is not able to fully reproduce the EW of emission lines in some cases (as seen in Figures \ref{fig:technical_seds_flex_vs_def} and \ref{fig:sedmodelling_chi2}), these incorrectly modelled emission lines are biasing the recent SFHs and derived physical properties in unknown ways. Given that emission line strengths are highly sensitive to ISM conditions that are not yet fully understood at high redshift \citep[e.g. temperature and density conditions, chemical abundances, ionizing spectra of stellar populations at high redshift etc., see][]{Reddy2023, Cameron2026}, by adding more flexibility in the emission lines, we avoid forcing the code to find a fit based on a range of parameters that may not be representative of the ISM in these galaxies.

Our solution for the version of the DJA SED catalogue used in this paper is therefore to have our fits driven solely by the richness in the shape of the stellar continua of these spectra. As will be discussed further on in this paper, we nevertheless recover key physical properties like SFRs on short timescales and dust attenuations. For the catalogue release, we make both versions of the catalogues available (flexible and default, see Section \ref{subsec:catalogue_release}) and leave it up to the user to choose their preferred version.

\subsection{{\sc Djapipes}}
\label{subsec:djapipes}

The SED fitting procedure described above has been implemented in a {\sc Python} wrapper called {\sc Djapipes}. This wrapper runs our custom version of \bagpipes{} under the hood and comprises several helper functions used to e.g., initialise the calibration polynomial as described in Section \ref{subsec:phot_calibration}, implement the emission line marginalisation as described in Section \ref{subsec:line_marginalisation}, mask emission lines if required, etc. Crucially, this wrapper interfaces with the spectroscopic and photometric databases on the Dawn \jwst Archive, thus simplifying the spectroscopic modelling of any PRISM spectrum with \jwst wide band photometry.

{\sc Djapipes} also includes post-processing capabilities such as custom plots for the SED fits and star-formation histories, posterior catalogues which contain the standard SED fitting outputs as well as emission line fluxes, posterior calibration curves, and helper functions to calculate quantities such as timescales of stellar mass build-up (e.g. $t_{10}$, $t_{50}$, $t_{90}$), SFR timescales, UV properties such as $\beta_{UV}$ and $M_{UV}$ and the strength of the Balmer and 4000 \AA{} breaks. {\sc Djapipes} is publicly available on GitHub\footnote{\protect \url{https://github.com/RashmiGot/dja-pipes}}. The database of our modelling results is accessible on the DJA. As a result of uniformly applying our modelling method to all {\tt grade=3} NIRSpec/PRISM spectra and corresponding wide-band NIRCam photometry, we note that within this selection lie lensed sources, LRDs, BL-AGN, transients, and other possible sources that this SED modelling set-up is not optimised to model. While we make all our modelling results available, we leave it to the discretion of the user to flag these sources when making use of the catalogue.

\section{Results}
\label{sec:results}

With our SED modelling implemented on thousands of PRISM spectra, we derive SFHs and physical parameters including stellar masses, star-formation rates on multiple timescales, and $V$-band dust attenuations. In the following sections, using the results from the flexible line modelling framework described above, we give an overview of these properties for our sample. We provide summary statistics of the physical properties (Section \ref{subsec:physical-properties}), summarise the star-formation rates that our fitting recovers, explore colour-colour classifications of galaxies through \ugis{} diagrams \citep[][]{Danso2023}, and look into the physical properties of the most dusty and quiescent galaxies in our sample (Sections \ref{subsec:dusty_galaxies}).

Examples of SED fits and SFHs for a number of galaxies are shown in Figure \ref{fig:sedmodelling_specfits}, ranging from cosmic noon to cosmic dawn ($z=2-14.4$). We model a diverse range of spectra, from galaxies at the highest observable redshifts, which are dominated by young stellar populations and strong UV lines, to intermediate- and lower-redshift sources encompassing strongly star-forming, extremely dusty, post-starburst, and quiescent galaxies. Importantly, we recover a diverse range of SFHs in this sample, with SFH shape and median stellar age largely independent of the observed redshift. This further motivates a census of the physical properties of these galaxies across cosmic time.

\subsection{Physical properties of galaxies at \texorpdfstring{$z_{\rm spec}>2$}{Lg}}
\label{subsec:physical-properties}

The distribution of stellar mass as a function of spectroscopic redshift is shown in Figure \ref{fig:results-zspec-vs-mass}. Our sample spans a stellar mass range of $10^6-10^{12}\ M_\odot$. The distribution peaks at ${\rm 10^{8.6}\ M_\odot}$, with a majority of our sample being low-mass galaxies. The shaded region above ${\rm 10^{10}\ M_\odot}$ indicates the threshold above which we define massive galaxies to reside. The number of massive galaxies peaks at $z\sim3$ and rapidly declines at $z\gtrsim6$, consistent with the expected decrease in the space density of massive galaxies at early cosmic times \citep{Weaver2022a, Weibel2024_smf, Shuntov2025}, compounded by the increasing difficulty of spectroscopically confirming such sources at these redshifts \citep{Danso2023}.

Figure \ref{fig:results-zspec-vs-mass} also shows surviving stellar mass as a function of rest-frame \ug{} colour, with each point coloured by its H$\alpha$+[N\,{\sc ii}] EW (calculated using the flexible line methodology; see Appendix \ref{appendix:ha_flux_eqw}). As galaxies increase in stellar mass, their stellar continua redden, as evidenced by their increasing \ug{} colour and an overall decrease in their H$\alpha$+[N\,{\sc ii}] EWs, possibly due to predominantly ageing stellar populations \citep{Whitaker2012, Whitaker2013, Belli2017}. The most massive galaxies above ${\rm 10^{10}\ M_\odot}$ have overall lower H$\alpha$+[N\,{\sc ii}] EWs and a broader range of \ug{} colours, reflecting the diversity of the massive galaxy population, including unobscured star-forming galaxies (with smaller Balmer breaks and bluer \ug{} colours), as well as dusty, star-forming and  post-starburst/quiescent galaxies (with redder \ug{} colours predominantly due to dust or age, respectively, with the latter population having intrinsically larger Balmer breaks).

\begin{figure*}[htb!]
\includegraphics[width=0.7\linewidth]{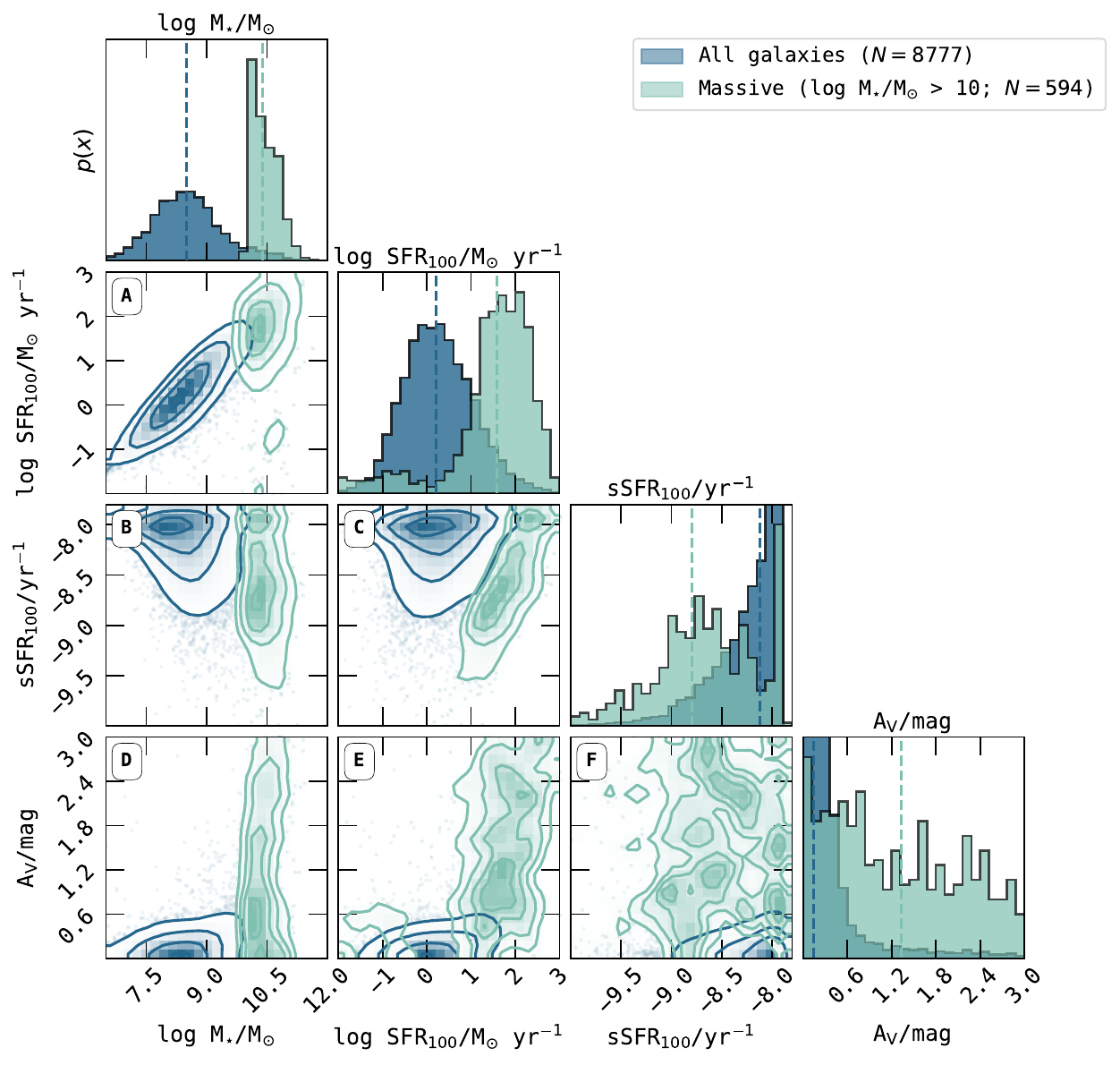}
\caption{
Corner plot of the median values of the posteriors of key physical properties of the DJA PRISM sample. Blue shows the full sample and green shows the massive galaxies. The dashed lines indicate the medians of the distributions.
The massive galaxies in our sample on average have higher SFRs, a broader range of sSFRs, and higher \av{}.
}
\label{fig:phys_props_hist}
\end{figure*}

The distributions of key physical properties for this sample are shown in Figure \ref{fig:phys_props_hist}.
Panel A shows that the overall galaxy sample is dominated by lower mass galaxies, and that the massive galaxies also represent the most highly star-forming ones. Panels B and C show that while the full sample uniformly spans a wide range of sSFR parameter space, the massive galaxies roughly fall into a high-sSFR clump and a low-sSFR clump.

Panels D, E and F show how the dust attenuation varies with mass, SFR and sSFR. The full sample has generally low \av{}, below 0.5 mag. The massive galaxies show clearer correlations with dust attenuation: \av{} shows a broader range at higher masses (panel D) and higher star-formation rates (panel E), hinting that dusty galaxies occupy the high mass end of the galaxy star-forming main-sequence (as discussed further in Section \ref{subsec:dusty_galaxies}).

From these derived properties, a picture emerges: the full spectroscopic DJA sample is dominated by low-mass, highly star-forming galaxies, a majority of which have very low dust attenuation. The most massive galaxies, those in the parent sample with stellar masses above ${\rm 10^{10}\ M_\odot}$, are organised in two clumps -- high and low sSFRs. Massive galaxies also show a strong correlation between stellar mass and SFR with rising \av{} \citep{Whitaker2017}, showing that the majority of massive galaxies lie in a distinct section of \av{} parameter space.

\subsection{Star-formation rates on 100, 10 and 3 Myr timescales}
\label{subsec:sfrs}

A variety of empirical indicators trace star-formation activity associated with different physical processes and characteristic timescales.  
The UV stellar continuum and far-IR dust continuum trace star formation on timescales of 100 Myrs and longer, while emission-line tracers such as $H\alpha$ and other recombination lines, originating in the H\,{\sc ii} regions ionised by the youngest, most massive stars, are sensitive to star formation on timescales of a few to $\sim$20 Myrs \citep{Kennicutt1998-review}. These classical indicators each calibrate a single characteristic timescale, typically assuming a constant or near-constant SFH; full spectrophotometric SED fitting with a flexible SFH, by contrast, recovers the SFR as a continuous function of lookback time directly from the fit, in principle allowing SFR estimates on any timescale without relying on separate tracer-specific calibrations.  Given our spectroscopic dataset, with NIRSpec/PRISM wavelength coverage from ${\rm 0.5-5.5\ \mu m}$ using the latest DJA reductions \citep{Brammer2025_DJA-spec}, we have access to the rest-frame UV continuum of galaxies at $z>2$, and to a host of emission lines such as H$\alpha$ up to $z\sim7$ and H$\beta$ up to $z\sim10$. In this section, we discuss the sensitivity of our SED fits to SFRs on these various timescales. Further, since the results presented here are from fits that are agnostic to the presence of emission lines (see Section \ref{subsec:line_marginalisation}), we discuss SFR estimates in the context of the H$\alpha$+[N\,{\sc ii}] EW to determine the sensitivity of the fits to shorter timescale star formation.

\begin{figure*}[htb!]
\includegraphics[width=\linewidth]{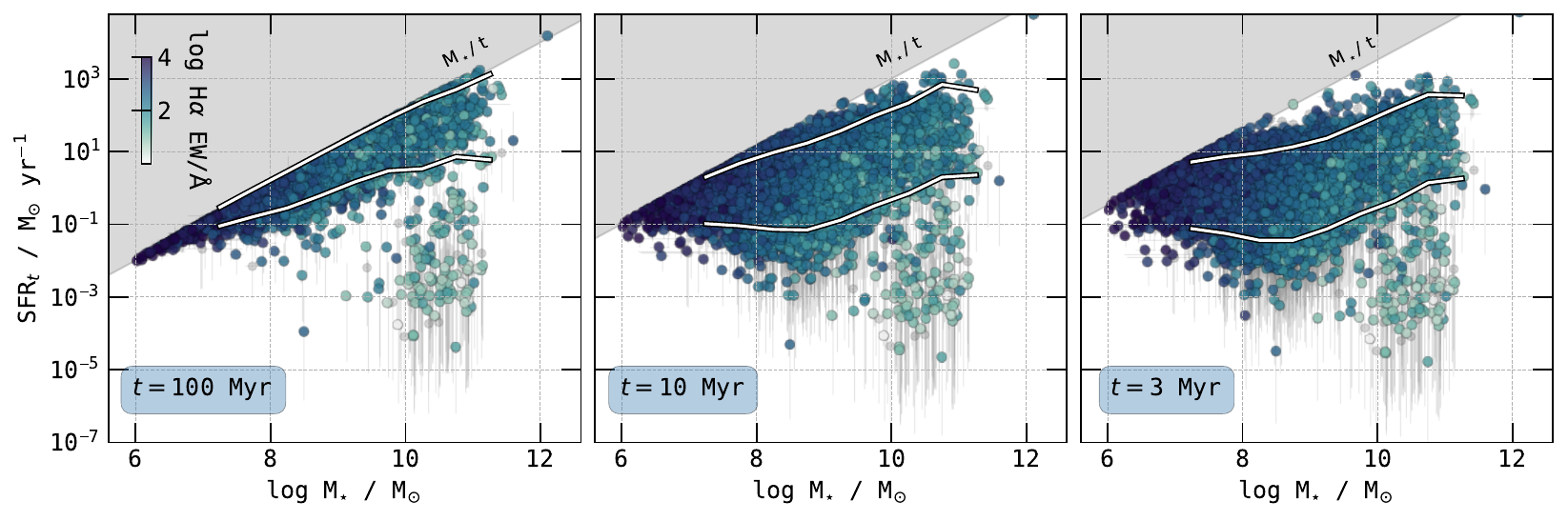}
\caption{Galaxies on the ${\rm M_\star-SFR}$ plane, coloured by the H$\alpha$+[N\,{\sc ii}] EW. On the left, the SFR is averaged over 100 Myr, in the centre over 10 Myr and on the right over 3 Myr. The two white lines encompass the 90th percentile range of SFR$_{t}$ in bins of 0.5 dex in stellar mass. The shaded region of the diagram shows the maximum possible SFR a galaxy of stellar mass ${\rm M_\star}$ can have, given the SFH averaging timescale.}
\label{fig:ms_ha_vs_ssfr_diff_timescales}
\end{figure*}

The location of our sources on the ${\rm M_\star-SFR}$ plane is shown in Figure \ref{fig:ms_ha_vs_ssfr_diff_timescales}, coloured by H$\alpha$+[N\,{\sc ii}] EW. We note the apparently tight SFR ceiling visible in all three panels of Figure \ref{fig:ms_ha_vs_ssfr_diff_timescales}. This is an artifact of the timescales over which our SFHs are averaged in order to derive SFRs. The upper limit for ${\rm SFR}_t$ (where ${t}$ is the averaging timescale, e.g. 3, 10 or 100 Myrs) is set by a galaxy of any given mass ${\rm M_\star}$ entirely forming within the last $t$ Myrs, such that ${\rm SFR_{t,max} = M_\star / t}$. E.g., a $10^8\ {\rm M_\odot}$ galaxy can have a maximum ${\rm SFR_{100}}$ of ${\rm 1\ M_\odot/yr}$ and a maximum ${\rm SFR_{10}}$ of ${\rm 10\ M_\odot/yr}$.

We quantify the typical width of the SFR distribution to be the average 90th percentile range of SFR in bins of 0.5 dex of stellar mass. For SFRs averaged over 100, 10, and 3 Myr timescales, the typical scatter in SFR is 1.36, 2.2, and 2.25 dex respectively (galaxies with ${\rm \log sSFR_{100}/yr < -11}$ are excluded from this calculation). The scatter in SFRs is higher on the 3 and 10 Myr timescales in comparison with the 100 Myr timescale, which has a much tighter ${\rm M_\star-SFR}$ relation. This follows from the fact that an overall rising SFH will have higher SFRs on shorter timescales, while an overall falling SFH will have lower SFRs on those same short timescales. This scatter is therefore a reflection of the diversity in the shapes of the recent star-formation histories in our sample. A similar increase in the scatter of the ${\rm M_\star-SFR}$ relation was shown in \cite{Cole2025} with a photometric sample up to $z_{\rm phot}=12$ and in \cite{Simmonds2025} with a sample up to $z_{\rm phot}=9$.

\begin{figure*}[htb!]
\includegraphics[width=\linewidth]{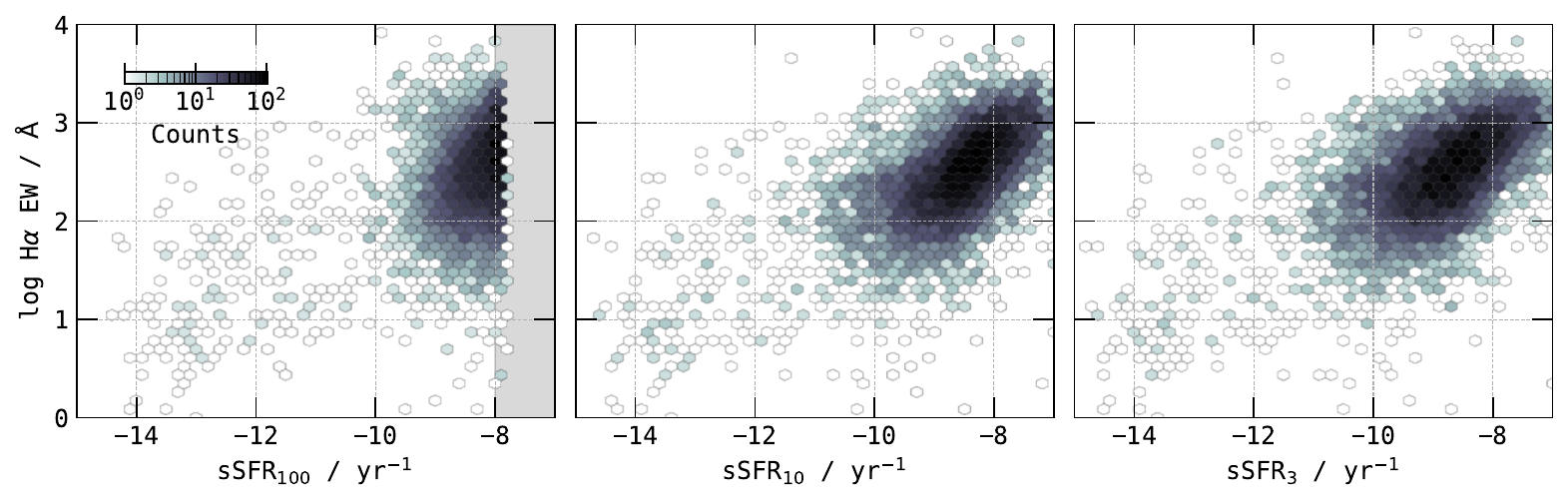}
\caption{H$\alpha$+[N\,{\sc ii}] EW vs. specific SFR calculated on a 100 Myr timescale (left), 10 Myr timescale (centre) and 3 Myr timescale (right). As shown, the H$\alpha$+[N\,{\sc ii}] EW is more tightly correlated with the sSFR on 3 and 10 Myr timescales than with the 100 Myr timescale.
}
\label{fig:ha_vs_ssfr_diff_timescales}
\end{figure*}

\begin{figure*}[htb!]
\includegraphics[width=\linewidth]{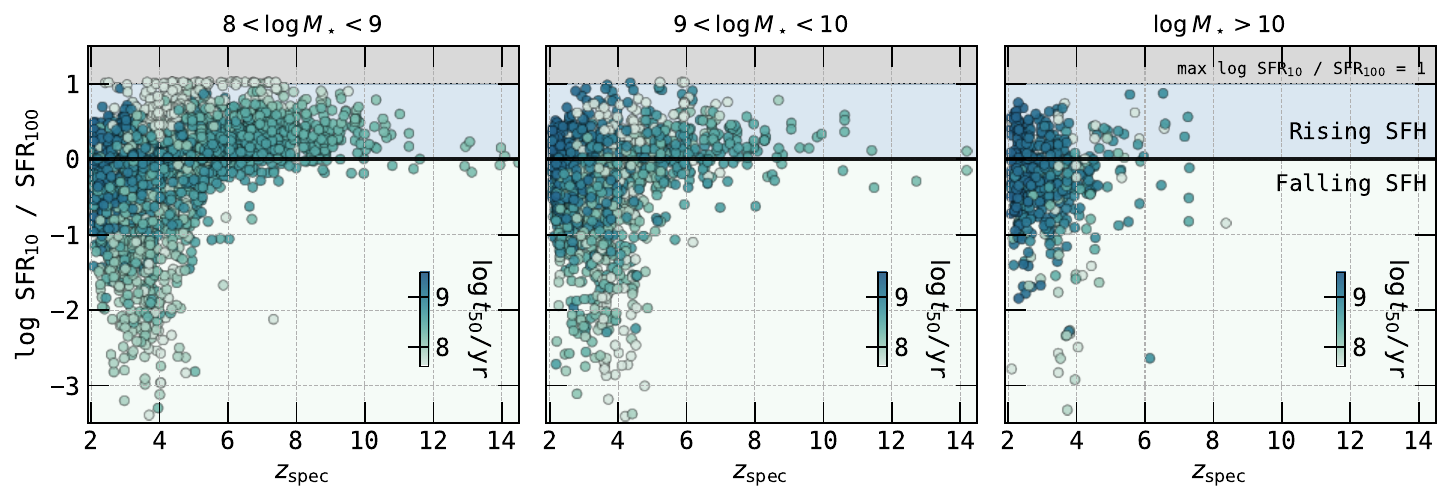}
\caption{Logarithmic ratio of \sfrt\ over \sfrh\ as a function of spectroscopic redshift for galaxies with ${\rm 8< \log M_\star / M_\odot < 9}$ (left), ${\rm 9 < \log M_\star / M_\odot < 10}$ (centre) and ${\rm \log M_\star / M_\odot > 10}$ (right). ${\rm \log SFR_{10} / SFR_{100} > 0}$ denotes a rising SFH and ${\rm \log SFR_{10} / SFR_{100} < 0}$ denotes a falling SFH in the last 100 Myr. Each scatter point is coloured by the $t_{50}$ timescales which is a measure of the median stellar age. At $z>6$ a majority of sources in this sample have rising SFHs, while falling SFHs tend of appear at $z<6$. Galaxies with high $t_{50}$ values tend for fall within a range of ${\rm -1 < \log SFR_{10} / SFR_{100} < 1}$, while galaxies with the lowest $t_{50}$ tend to have rising SFHs. Interestingly at ${\rm \log SFR_{10} / SFR_{100} < -1}$ which shows steeply falling SFHs we find galaxies with low $t_{50}$, which implies that these are younger galaxies than are experiencing a recent downturn in star-formation.}
\label{fig:dja_nirspec_sfr10_sfr100_vs_redshift}
\end{figure*}

Figure \ref{fig:ms_ha_vs_ssfr_diff_timescales} also explores trends with the EW of the H$\alpha$+[N\,{\sc ii}] line complex. While no correlation with EW can be seen for star-forming galaxies on the ${\rm SFR_{100}}$ panel, a clear correlation with EW is visible with ${\rm SFR_{10}}$ and ${\rm SFR_{3}}$. The sources with the highest EWs have the highest SFRs at any given mass, with the direction of steepest change in EW being perpendicular to the SFR upper ceiling. This shows the sensitivity of our fits to star-formation rates on timescales less than 20 Myrs.

This sensitivity to timescales is further explored in Figure \ref{fig:ha_vs_ssfr_diff_timescales}, where the $H\alpha$ EW is plotted against the specific SFR of our sources on timescales of 100 Myr, 10 Myr and 3 Myr respectively. While no correlation is seen between H$\alpha$+[N\,{\sc ii}] EW and ${\rm sSFR_{100}}$, the central and right panels show a clear correlation between rising $H\alpha$ EW and rising ${\rm sSFR_{10}}$ and ${\rm sSFR_{3}}$. Since our fits in this version of the catalogue are agnostic to the presence of emission lines, this correlation is non-trivial. It implies that there is indeed information in the shape of the stellar continuum that drives the fits of these high EW sources towards rising SFHs.

\subsection{The recent SFHs of galaxies}
\label{subsec:sfh_shapes}

Since our fitting recovers star-formation rates on timescales down to 3 Myr, we next look into the shapes of the star-formation histories in the most recent 100 Myr. We choose a simple parametrisation of the recent SFH shape, ${\rm \log SFR_{10} / SFR_{100}}$: if this quantity is greater than 0, then ${\rm SFR_{10} > SFR_{100}}$ and we interpret this as a rise in star-formation rate over the last 100 Myr; if this quantity is less than 0, then ${\rm SFR_{10} < SFR_{100}}$ and we interpret this as a fall in star-formation rate over the last 100 Myr \citep[several works such as][have used the same or similar parametrisations to explore SFR stochasticity]{Cole2025, Carvajal-Bohorquez2025}. If ${\rm SFR_{10}}$ is comparable to ${\rm SFR_{100}}$, i.e. ${\rm \log SFR_{10} / SFR_{100} \approx 0}$ then this is either a flat recent SFH of a quenched source with no star-formation in the past 100 Myr. ${\rm \log SFR_{10} / SFR_{100}}$ has a natural maximum of 1, which corresponds to the steepest possible rise in SFH in this parametrisation where a galaxy formed entirely within the most recent 10 Myr.

Figure \ref{fig:dja_nirspec_sfr10_sfr100_vs_redshift} shows ${\rm \log SFR_{10} / SFR_{100}}$ as a function of spectroscopic redshift from $2< z< 14.4$ coloured by $t_{50}$, the median stellar age and the time since a galaxy formed 50\% of its stellar mass.
At $z>6$, most galaxies in our sample have ${\rm \log SFR_{10} / SFR_{100}} > 0$ and so have rising SFHs; these galaxies also tend to have formed most of their mass more recently than other parts of the galaxy population. Galaxies that have high $t_{50}$ values, i.e. those that formed early on, tend toward flatter SFHs in the last 100 Myr, while those with the lowest $t_{50}$ (the youngest galaxies) correspond to the most steeply rising or steeply falling recent SFHs. The massive galaxies in this sample largely have continuous recent SFHs with high median stellar ages.

Interestingly, while sources that lie in the range $-1 < \log(\mathrm{SFR_{10}/SFR_{100}}) < 0$ seem to have relatively high $t_{50}$ values, at ${\rm \log SFR_{10} / SFR_{100} < -1}$ there is a reversing gradient towards lower $t_{50}$ (younger ages) again. These are sources that have recently had a steep downturn in star-formation and have only recently formed a majority of their stellar mass; these could be objects that have recently quenched \citep[so-called ``mini-quenched'' galaxies, e.g.][]{Looser2023, Baker2025_miniquenched} or are simply going through a lull of star-formation \citep[so-called lulling or smouldering galaxies, e.g.][]{CoveloPaz2026, Trussler2025, Baker2026, Mintz2026}. These sources are most common in the lower mass bins of ${\rm 8< \log M_\star / M_\odot < 9}$ and ${\rm 9 < \log M_\star / M_\odot < 10}$.

The overall age gradient in Figure \ref{fig:dja_nirspec_sfr10_sfr100_vs_redshift} seems to suggest that while galaxies that formed longer ago (high $t_{50}$) have flatter recent SFHs, more recently formed galaxies (low $t_{50}$) fall into two categories: (a) those with steeply rising SFHs and (b) those with rapidly falling SFHs. Therefore younger sources appear to have more extreme gradients in star-formation while older galaxies have flatter recent SFHs. Overall, the trend with mass shows that massive galaxies tend towards flatter recent SFHs while lower mass galaxies have a diverse range of SFH shapes. A similar trend with mass was explored in \cite{Simmonds2025}, with a photometric sample up to $z_{\rm phot}=9$, showing that lower mass galaxies experienced increased ``burstiness'' in the most recent 50 Myr; we see a similar increase in scatter towards lower masses in the most recent 100 Myr.

The trend seen with redshift suggests that at $z\gtrsim6$, the SFHs are predominantly rising, while falling SFHs are seen more ubiquitously at $z\lesssim6$. This could represent an underlying transition in the galaxy population at $z\approx6$; this could also reflect the observational effect of having access to the rest-frame UV at $z\gtrsim6$ and thus being more sensitive to young stellar populations above this redshift. These redshift trends remain to be investigated with an unbiased observational sample.

\begin{figure*}[htb!]
\includegraphics[width=0.76\linewidth]{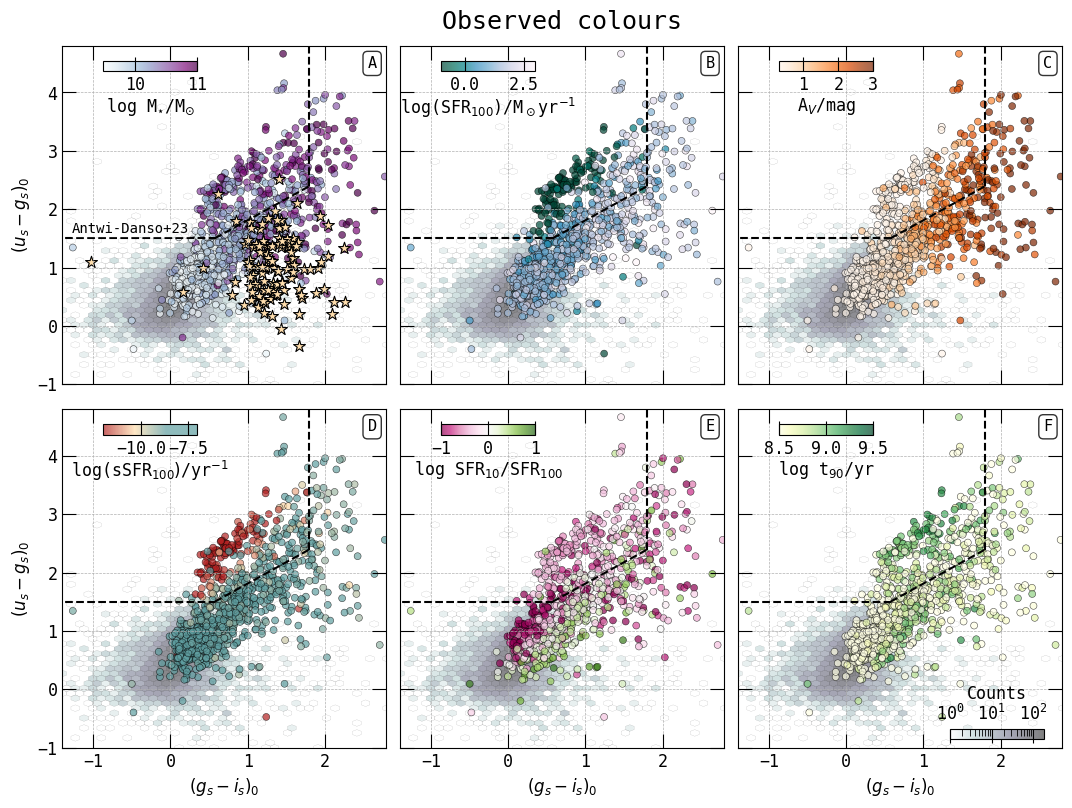}
\includegraphics[width=0.76\linewidth]{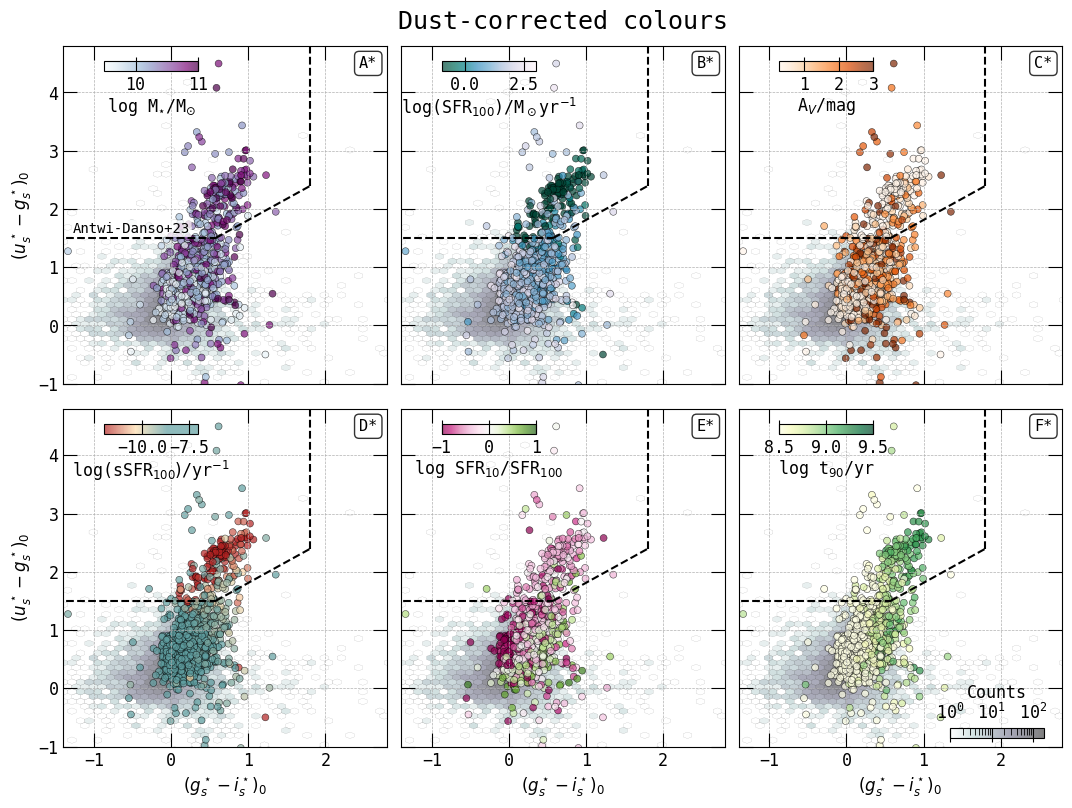}
\caption{Rest-frame and dust-corrected \ugis{} colour-colour diagrams of our sample of galaxies; coloured scatter points correspond to sources with ${\rm \log M_\star / M_\odot > 9.5}$, and lower mass sources can be seen in the background greyscale plot. Stars denote LRDs identified from \cite{deGraaff2025_LRDoverview}. From top left to bottom right, the observed (and below, the dust-corrected) \ugis{} diagram is coloured by A (A*): stellar mass, B (B*): star-formation rate, C (C*): V-band dust attenuation, D (D*): specific star-formation rate, E (E*): logarithm of the ratio of ${\rm SFR_{10} / SFR_{100}}$, and F (F*): $t_{90}$.}
\label{fig:ugi_sed_cbar}
\end{figure*}

\subsection{Physical properties of massive galaxies on colour-colour diagrams}
\label{subsec:colour_colour_diagrams}

Colour-colour diagrams are an efficient tool to explore of physical properties of massive galaxies.
Classifications such as the $UVJ$ have been extensively used to identify massive quiescent galaxies from photometric surveys \citep{Wuyts2007, Williams2009, Whitaker2011}, to establish a bimodal sequence of blue and red galaxies \citep{Brammer2009}, and to identify pathways for quenching \citep{Belli2019, Park2023}. The $(U-V)_{\rm rest}$ colour traces the strength of the Balmer and 4000 \AA{} break, and the $(V-J)_{\rm rest}$ colour traces the dust-reddening of the stellar continuum redward of the Balmer break, such that the $UVJ$ diagram efficiently separates young star-forming and dusty galaxies from old quiescent galaxies.  However, the utility of the $UVJ$ diagram at $z>3$ breaks down for several reasons \citep[see discussion in][]{Whitaker2026}.

Newer colour-colour classifications have been introduced to alleviate the issues with the $UVJ$ selection method at higher redshifts, such as emission-line contamination, dusty interlopers, and AGN at the red end, and young recently quenched galaxies at the blue end \citep[e.g.][]{Danso2023, Gould2023, Baker2025}. Here, we explore the distributions of various derived physical properties on \ugis{} colour-colour diagrams \citep[][we briefly discuss other $UVJ$ and \ugis{} selection criteria in Appendix \ref{appendix:uvj_ugi_selections}]{Danso2023}. We note that the \us{}, \gs{} and \is{} synthetic colours have been directly measured from the photometrically calibrated NIRSpec/PRISM spectrum by integrating the spectrum through the synthetic filter passbands, as opposed to the best-fit models.; we therefore do not extrapolate colours from the SED model and the data has a natural cut-off at $z=6.7$ where the \is{} band redshifts out of the observed wavelength range of NIRSpec/PRISM. We also show dust-corrected versions of the diagrams (which allow us to interpret physical quantities in the context of the intrinsic colours of the SED), for which the observed \us{}, \gs{} and \is{} magnitudes have been measured from the intrinsic, dust-corrected spectrum. We describe the exact dust-correction procedure in Appendix \ref{appendix:dust_correction}.

In Figure \ref{fig:ugi_sed_cbar}, we show the massive galaxies in our spectroscopic sample (${\rm \log M_\star / M_\odot > 9.5}$) scattered on the \ugis{} diagram, with each column showing the sample coloured by a different physical property derived from our SED modelling. The top panel shows the observed \ug{} and \gi{} colours, while the bottom panel shows the dust-corrected versions of the diagrams.

\begin{figure*}[htb!]
\includegraphics[width=\linewidth]{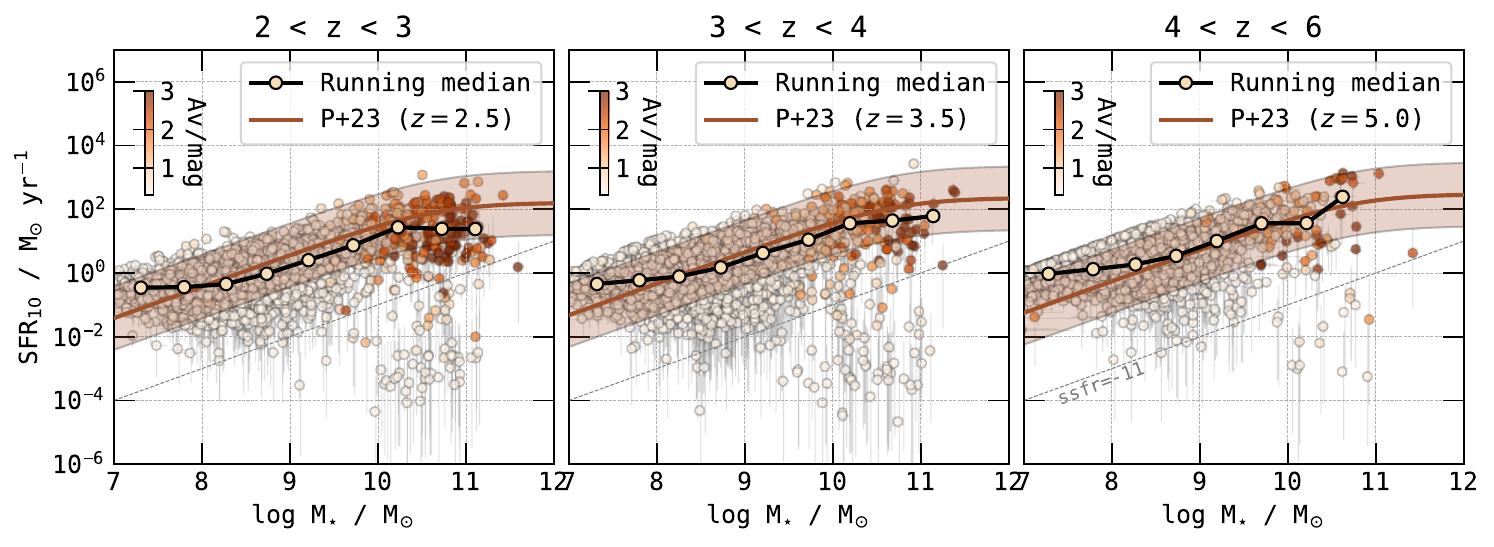}
\caption{The galaxy star-forming main-sequence at redshifts of $2<z<3$, $3<z<4$ and $4<z<6$. Galaxies are shown by scatter points and coloured by \av{}.  The star-forming main-sequence at $z=2.5,3.5,5.0$ from \cite{Popesso2023} is shown as the solid black line with the shaded region indicating $\pm1$ dex. Dusty galaxies on average populate the high-mass end of the galaxy main-sequence across all redshifts shown here.}
\label{fig:dusty_galaxies_MS}
\end{figure*}

Panel A (A*) of Figure \ref{fig:ugi_sed_cbar} shows a gradient in stellar mass, increasing towards the upper right of the \ugis{} diagram, with galaxies in the \citep{Danso2023} selection box corresponding to the most massive galaxies. Galaxies that lie beyond the red-most boundary of the box towards the upper right (which correspond to the dustiest sources; see Panel C (C*) of Figure \ref{fig:ugi_sed_cbar}, discussed below) also tend to be amongst the most massive galaxies in this sample.

In Panel B (B*), the gradient in ${\rm SFR_{100}}$ shows that quiescent galaxies (those with the lowest SFRs) form a tight sequence within the \ugis{} box. At a given \gi{} colour, the reddest \ug{} colours correspond to those sources with the lowest SFRs. While the tight sSFR sequence can be seen well into the interior of the \ugis{} selection box, a number of sources with non-negligible SFRs reside near the boundary of the \cite{Danso2023} selection, suggesting that these could be younger, recently quenched objects or dusty star-forming interlopers.

Panel C (C*) shows that at fixed \ug{}, redder \gi{} colours correspond to sources with higher \av{} which directly reflects the reddening of the optical stellar continuum due to dust. This is further seen in the dust-corrected version of the diagram, where the scatter towards the upper right of the diagram is ``corrected for'', showing the exact direction in which the dust vector acts. While galaxies in the interiors of the quiescent box are largely dust-free, galaxies on the boundary do contain significant amounts of dust.

In Panel D (D*), quiescent galaxies form a tight sequence within the \ugis{} box with the lowest sSFRs, simialr to the trend seen in Panel B (B*). In D*, the bluest \ug{} and \gi{} intrinsic colours correspond to the highest sSFRs, and the reddest \ug{} and \gi{} intrinsic colours correspond to those sources with the lowest sSFRs. Galaxies with low sSFRs are well captured by the \ugis{} selection of \cite{Danso2023}.

We explore gradients in the recent SFHs of galaxies in Panel E (E*) of Figure \ref{fig:ugi_sed_cbar}, parametrised by ${\rm \log SFR_{10} / SFR_{100}}$ (see Section \ref{subsec:sfh_shapes} for a discussion of this parameter). Positive values (green) indicate rising SFHs with ${\rm SFR_{10} > SFR_{100}}$ and negative values (purple) indicate falling SFHs in the last 100 Myr (as in Section \ref{subsec:sfh_shapes}). While sources with rising SFHs squarely reside in the star-forming region of the diagram, those with falling SFHs lie outside the lower left boundary of the quiescent box. These points represent galaxies with falling SFHs and therefore include sources that have turned off their star-formation within the last 100 Myr (or that are in the process of doing so). The fact that they lie outside or on the boundary of the quiescent selection box is because these sources host younger stellar population (on account of them having undergone an episode of star-formation at least 100 Myr ago) which make their \ug{} colours lower due to weaker Balmer breaks and the \gi{} colours lower due to bluer rest-optical stellar continua \citep[see][for a similar discussion in the context of the $UVJ$ diagram]{Baker2024_quiescent_SFHs}. These galaxies with falling SFHs likely represent a transitory population between star-forming and quiescent galaxies.

\begin{figure*}[htb!]
\includegraphics[width=\linewidth]{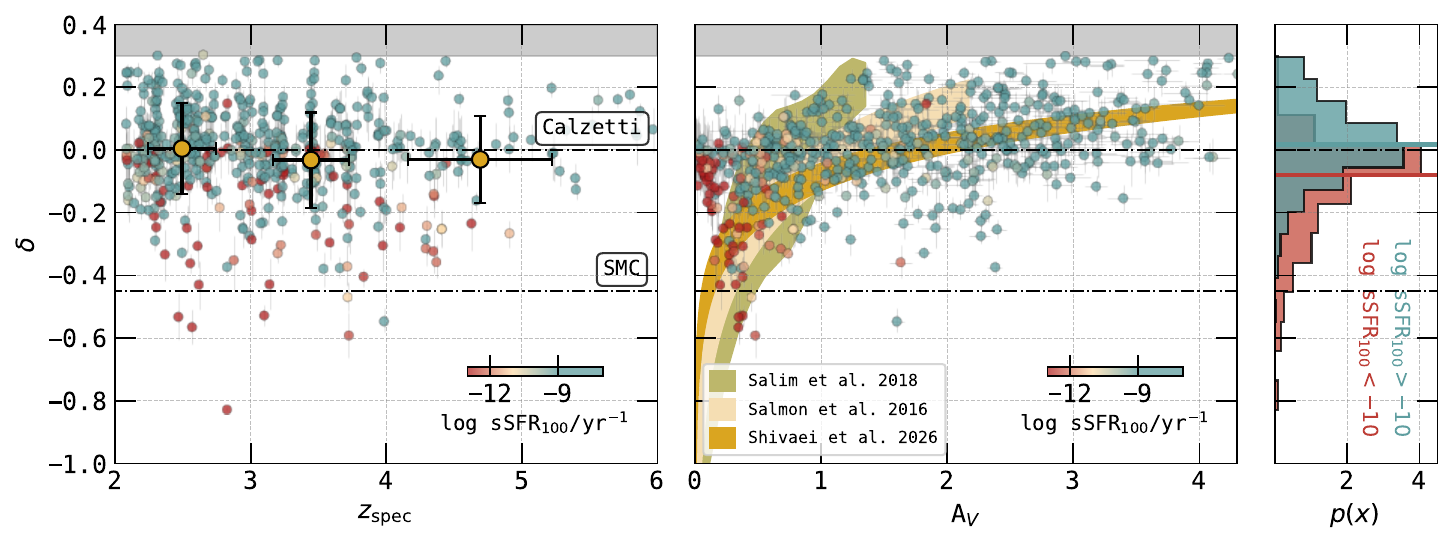}
\caption{$\delta$, the slope of the attenuation curve, plotted against spectroscopic redshift (left) and \av, the $V$-band dust attenuation (right) for the massive galaxies in our sample (${\rm \log M_\star / M_\odot > 10}$). Scatter points are coloured by the specific SFR. Histograms of $\delta$ are shown to the right-most panel and are split by ${\rm sSFR < -10}$ (red) and ${\rm sSFR > -10}$ (blue). Galaxies with low ${\rm sSFR < -10}$ and low \av s have steeper attenuation curves while galaxies with higher \av s have shallower attenuation curves in broader agreement with the local \cite{Calzetti2000} curve. $\delta-$\av{} relations from \cite{Salim2018_dust_law} (local universe), \cite{Salmon2016} ($z \sim 1.5 - 3$) and \cite{Shivaei2025} ($z \sim 1-9$) are shown; our sample is in broad agreement with these relations.\\
}
\label{fig:massive_galaxies_delta_Av}
\end{figure*}

Somewhat counter-intuitively, from the trends seen in Panel E (E*) that the dustiest galaxies in this sample have largely flat or recently declining SFHs. We note that this could be a result of the flexible line modelling method, as shown through an example in Figure \ref{fig:technical_seds_flex_vs_def}.

The final Panel F (F*) of Figure \ref{fig:ugi_sed_cbar} shows the \ugis{} diagram coloured by the logarithmic $t_{90}$ timescale, which measures the time since 90\% of a galaxy's stellar mass formed. $t_{90}$ traces the last major episode of star-formation, and so it follows that the older a galaxy is, the higher its $t_{90}$ value. While $t_{90}$ shows significant scatter in the observed \ugis{} colours, a much clearer picture can be seen in the dust-corrected version of the diagram: the galaxies with the reddest intrinsic \ug{} and \gi{} colours, i.e. those with the oldest stellar populations, do indeed have the highest $t_{90}$ values. Interestingly, we see two diagonal sequences entering the quiescent box, one with $\log t_{90} / {\rm yr} < 9.0$ and another with $\log t_{90} / {\rm yr} \gtrsim 9.0$: younger quiescent galaxies occupy the lower left region of the box while older quiescent galaxies occupy the upper right interiors of the box. This is suggestive of two distinct quenching tracks (discussed in Section \ref{subsec:quenching_pathways}).

Overall, galaxies with ${\rm \log M_\star / M_\odot > 9.5}$ in this spectroscopic sample range from quiescent to star-forming to dusty galaxies, and their properties in \ugis{} colour spaces vary continuously across traditional colour-selection thresholds. In the past, colour selections have been required to separate massive galaxies into distinct classes based on photometric colours, but the spectroscopically-derived trends here show that massive galaxies have continuously varying properties across the boundaries of the \ugis{} selection box. By avoiding classifications based on observed colours, we even find interesting transitory populations such as those showing a recent downturn in star-formation. Here, we emphasize the importance of studying the massive galaxy population as a whole rather than pre-selecting quiescent/dusty galaxies purely based on colours, so as to catch all phases of massive galaxy evolution.

\subsection{The physical properties of massive dusty galaxies}
\label{subsec:dusty_galaxies}

In this section, we look at the main-sequence properties of galaxies in the context of their dust attenuation. Figure \ref{fig:dusty_galaxies_MS} shows the star-forming main sequence at redshifts of $2<z<3$, $3<z<4$ and $4<z<6$, coloured by \av{}. We note that the SFRs plotted here are ${\rm SFR_{10\ Myr}}$.

The dustiest galaxies occupy the high-mass end of the distribution and tend to fall predominantly within $\pm 1$ dex of the galaxy star-forming main-sequence relation, while a non-negligible fraction of dusty sources scatter below the main-sequence and towards the quiescent region. These findings are in line with studies of optically faint galaxies conducted with \hst{}, \jwst{}, and previously Spitzer, which showed that dusty galaxies tend to be massive \citep{Barrufet2024} and on the star-forming main-sequence \citep{Wang2019, Barrufet2023, Gottumukkala2024}.

In Figure \ref{fig:dusty_galaxies_MS}, while the number of dusty galaxies decreases towards higher redshifts, the dustiest sources continue to occupy the high mass end of the galaxy main-sequence, further supporting the tight coupling of stellar mass and dust build-up. The overall scarcity of dusty galaxies towards higher redshifts could be due to high-redshift galaxies not having enough cosmic time to form small dust grains. It could also likely be explained by the scarcity of massive galaxies in general at $z>5$, further supporting the coupling of mass and dust build-up. It could also be a result of primarily catching the rest-frame UV emission of galaxies at these high redshifts, which is more acutely plagued by extinguished light due to dust than the rest-frame optical \citep{Manning2025}.

Given our use of the flexible \cite{Salim2018_dust_law} dust law (albeit with a strong prior on the \cite{Calzetti2000} law), we explore the slope of the attenuation curves in Figure \ref{fig:massive_galaxies_delta_Av} for the massive galaxies in our sample (${\rm \log M_\star / M_\odot > 10}$). We show $\delta$, the deviation of the slope from that of the \citep{Calzetti2000} the attenuation curve (which captures how steep/shallow the attenuation curve) is vs. spectroscopic redshift and \av{}, where our sample displays a broad range of $\delta$ values at low \av{}, whereas the average value of $\delta$ increases with increasing \av{} for the sample overall. As galaxies become dustier, i.e. towards larger optical depths, their attenuation curves become greyer (shallower). This trend has been shown in e.g. \cite{Salmon2016, Salim2018_dust_law,Salim2020_dust_review, Shivaei2025, Chworowsky2026}, and potentially arises from dustier sources having more complex star-dust geometries such as gradients from face-on to edge-on inferred dust attenuation.

Several works have noted a weak redshift evolution in the $\delta-$\av{} relation, with higher redshift sources displaying shallower attenuation curves \citep{Shivaei2025, Chworowsky2026}. While we do not see a noticeable evolution with redshift, we cannot rule it out based on this sample. We note that the sources studied here are restricted to the massive end, so we may be missing a majority of higher redshift sources studies in e.g. \cite{Shivaei2025}. Instead, we find a stronger correlation with ${\rm sSFR_{100}}$; for low sSFRs, the \av{}s are lower relative to the rest of the sample and the $\delta$ values span a wide range but importantly scatter towards lower $\delta$ (i.e. steeper curves) than sources with higher dust attenuation. This shows that quiescent galaxies have steeper attenuation curves on average than dusty star-forming galaxies. This trend with quiescent galaxies having steeper attenuation curves was shown in \cite{Salim2018_dust_law} for a sample of local galaxies; here we show that this trend holds at high redshifts, beyond $z=2$. This is likely due to quiescent galaxies having lower dust columns and simpler, less clumpy star-dust geometries; as a result, the intrinsic steepness of the extinction curve is preserved, as opposed to being greyed out by scattering and saturation effects arising from higher dust columns and more complex geometries \citep[see a discussion in][]{Whitaker2026}

\section{Discussion}
\label{sec:discussion}

From a diversity of SEDs and SFH shapes to smoothly varying properties in colour-colour spaces, we have seen that this sample of spectroscopically observed high redshift galaxies cover a wide range of stellar masses, star-formation rates and dust attenuations. In the following sections, we discuss the utilisation of this catalogue and contextualise our findings within the broader picture of high-redshift galaxy evolution.

\subsection{Stochastic star-formation}
\label{subsec:bursty_sf}

Recent results from \jwst have widely discussed the bursty and stochastic nature of star-formation \citep[e.g.][]{Cole2025, Ciesla2024, Carvajal-Bohorquez2025, Simmonds2025} and its link with the UV excess of galaxies at high redshifts \citep[e.g.][]{Donnan2023, Harikane2023, McLeod2024}. Studies have found that low-mass galaxies at $z>6$ tend to have higher variability in their star formation on shorter timescales compared with higher mass systems at lower redshifts \citep{Simmonds2025,Looser2025}. This is expected at redshifts where galaxies are hosted by lower mass dark matter halos with shallower potential wells, which are more conducive to feedback driven outflows and therefore to stochastic star formation.

In Figure \ref{fig:ms_ha_vs_ssfr_diff_timescales}, the typical scatter in the ${\rm M_\star-SFR}$ plane increases from ${\rm SFR_{100} \to SFR_{10} \to SFR_{3}}$ by 1.36, 2.2, and 2.25 dex respectively, i.e. shorter SFR averaging timescales have a larger scatter in observed SFRs \citep[as also seen in e.g.][]{Cole2025, Carvajal-Bohorquez2025}. This trend is further explored as a function of redshift and stellar mass in Figure \ref{fig:dja_nirspec_sfr10_sfr100_vs_redshift}, where a majority of galaxies at $z>6$ display rising SFHs within the last 100 Myr, whereas significantly more diverse shapes are seen at $2<z<6$.

The fact that we do not find significant numbers of galaxies with recent falling SFHs at $z>6$ could very well be an effect of the spectroscopic target selections; it could be a consequence of several \jwst surveys targeting UV bright sources at the expense of missing galaxies that have lower ${\rm SFR_{10}}$ and therefore relatively less light from O- and B-type stars dominating their stellar continua. Relatedly, the lack of recent falling SFHs at $z>6$ could also be an observational effect: beyond $z\sim6$, access to the red-most part of the rest-frame optical becomes increasingly limited. This wavelength regime is most sensitive to older stellar populations and the bulk of the stellar mass of galaxies \citep{Papovich2023}. The SFHs at $z>6$ are thus increasingly constrained by the rest-frame UV emission, which is most sensitive to young stellar populations and could preferentially infer rising recent SFHs. Alternatively, it could also be a physical trend where galaxies at high redshifts tend to have steeply rising recent SFHs as a result of having higher gas fractions, and therefore more vigorous star-formation.

Figure \ref{fig:dja_nirspec_sfr10_sfr100_vs_redshift} suggests an alternate link of stochasticity driven mainly by stellar mass. The most massive galaxies in this sample with ${\rm \log M_\star / M_\odot > 10}$ tend towards flatter recent SFHs while the lower mass galaxies show a broader range of SFH shapes within the last 100 Myr. At $z>6$, since our sample largely consists of low-mass galaxies with ${\rm \log M_\star / M_\odot < 10}$, we are preferentially picking up sources with rising recent SFHs (and lose those with falling SFHs due to selection effects).

It is worth noting that, given our methodology of continuum-driven modelling that decouples the derived SFHs from the emission lines, our work is distinct from studies that use the H$\alpha$ to UV ratio as an indicator for stochasticity. Instead, the range of SFH shapes are purely coupled to the detailed information in the stellar continuum.

\subsection{Quenching in massive galaxies}
\label{subsec:quenching_pathways}

Amongst the most surprising results from \jwst have been discoveries of spectroscopically confirmed massive quiescent galaxies in the first 1.5 Gyr of cosmic history \citep{Weibel2024b, deGraaff2024a, Carnall2023, Glazebrook2024, Baker2024_quiescent_SFHs}. The process of formation and cessation of star-formation at high-redshifts must presumably be rapid, given the early periods of cosmic time in which these galaxies assembled. Studies have shown an increase in rapid vs slow quenching towards cosmic noon \citep{Belli2019, Park2023}, but this remains to be decisively seen into the first billion years. Additionally, given that the process of quenching unfolds on timescales of $\lesssim100$ Myr to a few gigayears, there must be a transitional population at these redshifts that are in between vigorous star-formation and complete quenching \citep{Fang2018, Whitaker2026}.

The suite of \ugis{} diagrams presented in Figure \ref{fig:ugi_sed_cbar} of massive galaxies (${\rm \log M_\star / M_\odot > 9.5}$) looks at various SFH-derived quantities to determine overall trends in mass assembly. Galaxies with the lowest sSFRs sit well within the \ugis{} colour selection box (Figure \ref{fig:ugi_sed_cbar}, Panel D*). To the lower left of this population sit galaxies experiencing a downturn or shutting off of their star-formation (Panel E*); with ageing and subsequent reddening of \ug{} and \gi{} colours, this population could quench and enter the quiescent region of the diagram. Taken together with $t_{90}$ (Panel F*), we see that this population is young relative to galaxies with redder \gi{} colours, and could potentially hint at a faster quenching track. This would be in line with findings in \citet{Xiao2026}, a study based on the catalogue presented here, that found distinct rapid and extended quenching pathways in the most massive galaxies in this sample.

Studies have shown that massive quiescent galaxies in the early universe show a wide variety of ages and quenching timescales \citep{Nanayakkara2024, Nanayakkara2024b, Baker2024_quiescent_SFHs}.
Connecting transitional populations to fully quenched populations and determining modes of quenching requires a more detailed study of the assembly histories of the massive galaxies in this catalogue.
The exact shapes of the SFHs of quiescent galaxies relative to the entire massive galaxy population in this sample remain to be studied in detail, and we defer this to future work.

\subsection{Dust properties of massive galaxies}
\label{subsec:dust_production}

The assembly of the first massive galaxies is closely coupled with the assembly of the first dusty galaxies. Early \jwst studies based on photometric selections detected massive, dusty galaxies up to $z\sim8$, showing their prevalence in the first 2 Gyr and their significant contribution to the obscured star-formation rate density and stellar mass functions \citep{Barrufet2023, Gonzalez2023, Gottumukkala2024, Weibel2024_smf, Williams2023}. In line with several of these studies, our work shows that the dustiest galaxies do indeed lie largely on the star-forming main sequence and represent the most massive end of the galaxy population in this sample. This implies a tight relationship between stellar mass and dust build-up in the first few Gyr of cosmic history.

Studies with NIRSpec spectroscopy have begun to reveal the stellar populations and detailed dust attenuation laws of massive dusty galaxies \citep{Cooper2025, Markov2025}. The general trend of greyer attenuation slopes as optical depth increases is seen in our sample and follows other works in the literature \citep{Shivaei2025, Chworowsky2026, Maheson2025}. Interestingly, we notice a trend with sSFR, where galaxies with lower sSFRs appear to have steeper attenuation slopes \citep[seen previously in local galaxies,][]{Salim2018_dust_law}. This could point towards different grain-size distributions or star-dust geometries associated with distinct populations of galaxies.

Several trends of the dusty galaxies within this population remain to be explored, e.g. the variation of the slope of the attenuation curve with stellar mass and age and the emergence of the $2175 \text{\AA}$ UV bump.

\subsection{Caveats}
\label{subsec:caveats}

The sample studied in this paper covers a broad range of observed redshifts from $z=2$ to $z=14.4$, corresponding to the first three billion years of cosmic history. As such, given the uniform observed wavelength coverage of the PRISM from ${\rm 0.5-5.5\ \mu m}$ for all sources, the rest-frame wavelengths probed in this sample range from ${\rm 0.16-1.8\ \mu m}$ at $z=2$ (covering the rest-frame UV -- NIR) to ${\rm 0.03-0.35\ \mu m}$ at $z=14.4$ (covering predominantly the rest-frame UV). The youngest stellar populations which consist of young O- and B-type stars dominate the emission in the rest-frame UV and at the blue end of the rest-frame optical, thus making the highest redshift objects in this sample particularly susceptible to outshining of older stellar populations. We note that this could likely drive some of the trends seen in this paper, e.g. finding predominantly rising SFHs at $z>6$. Spatially resolved studies are needed to determine the extent of the outshining problem, \citep{GimenezArteaga2023, Harvey2025, Mosleh2025, Wang2025, Narayanan2024} with IFU studies being of particular importance.

Our modelling results are intimately coupled with the underlying stellar spectral libraries used within \bagpipes{}. For this catalogue, we used the BPASS stellar libraries with the binary star prescription. Crucially, BPASS models do not include modelling of the thermally-pulsating asymptotic giant branch (TP-AGB) phase of stellar evolution, which can affect the light in the NIR by up to a few factors on timescales of 0.3 - 2 Gyr \citep{Lu2025_Nature, Conroy2013}. In an accompanying study \citep{Xiao2026}, we show that replacing the fiducial BPASS models with the \cite{Bruzual2003} (BC03) stellar population synthesis models has a negligible effect on the resulting stellar mass estimates for our sample at $z>3$ \citep[although we note that the BC03 models have negligible TP-AGB contribution to the overall SED; see e.g.][]{Lu2025_Nature}. The effects of changing the stellar spectral libraries on the overall modelling results in this paper, in particular on the star-formation histories, remain to be tested \citep[see e.g.][for similar analyses]{Nanayakkara2024b, Wang2024, Whitler2023}.

Also affecting our modelling results are the priors imposed on the SFHs themselves. Parametric SFH models impose strong priors on inferred physical properties \citep{Carnall2019b}, while non-parametric priors are capable of recovering a larger diversity of SFH shapes \citep{Leja2019}, thus motivating our use of the continuity prior. While stellar masses are fairly robust against various parametric and non-parametric priors \citep[e.g.,][]{Cochrane2025, Xiao2026}, the shapes of the SFHs can often differ between the two classes of priors \citep{Leja2019}. More recently, studies have found that physically-motivated SFH priors such as those informed by the dark-matter accretion rate and/or coupled with the star-forming main-sequence can alleviate tensions of overly massive or maximally old galaxies at high redshifts \citep[e.g.,][]{Turner2025, Duan2026}. While not yet tested in this paper, parametric and non-parametric SFH priors will be explored for this sample in an upcoming paper.

Recent studies with \jwst{} have increasing shown that galaxies at high redshifts tend to have non-solar metal abundance patterns \citep[see e.g.][]{Beverage2025, Carnall2024, Shapley2025}, a direct consequence of rapid galaxy formation and expected when studying stellar populations within the first billion years of cosmic evolution. Commonly used SPS libraries such as BPASS and BC03 typically only include solar-scaled abundances (i.e. the inferred metallicities implicitly assume that the relative elemental abundances are solar). This can bias key inferred quantities, such as metallicities, and therefore the SFHs of galaxies. Folding in $\alpha$-enhanced SPS models into the modelling framework, such as $\alpha$-MC \citep{Park2025}, will be crucial to determining the robustness of the SFHs and physical properties presented in this study, in particular for the massive galaxy population.

\subsection{Catalogue release}
\label{subsec:catalogue_release}

The catalogue presented here will be made available on Zenodo\footnote{\cite{Gottumukkala2026_Zenodo}; \href{https://zenodo.org/records/22734795}{zenodo.org/records/22734795}}. We provide results from our flexible line modelling methodology as well as the standard fitting method with \bagpipes{}. The following is an overview of the main functionalities provided:

\begin{itemize}
    \item Searchable database of SED fits, SFHs and physical properties for the full sample on the DJA\footnote{Online overview tables for  \href{https://msaexp-nirspec.s3.us-east-1.amazonaws.com/pipes/run_2606/release-v4/djapipes_stellar_pop_flexible_v4.6.0.html}{flexible} and \href{https://msaexp-nirspec.s3.us-east-1.amazonaws.com/pipes/run_2606/release-v4/djapipes_stellar_pop_default_v4.6.0.html}{default} fitting methods}
    \item Catalogues of physical properties, SED fits, SFHs, dust curves, calibration curves, line fluxes
    \item Jupyter Notebook with access details and examples scripts
\end{itemize}

An example table of some of the available data products are shown in Table \ref{tab:phys_props} of the appendix.

\section{Summary and conclusion}
\label{sec:conclusion}

In this paper, we present an archival catalogue of SED modelling properties of all public \jwst NIRSpec/PRISM spectra, released on the Dawn \jwst Archive. This paper presents the functionalities of this catalogue and gives an overview its main scientific applications.

\begin{itemize}

\item This sample comprises $\sim 12,000$ high-quality \& visually-vetted NIRSpec/PRISM spectra from $2<z<14.4$, drawn from v4.5 of the DJA. We jointly model the PRISM spectra and NIRCam wide band photometry of these sources with a custom version of the SED modelling code \bagpipes{} \citep{Carnall2018, Carnall2019} with a flexible line modelling technique, building a catalogue of inferred SFHs and stellar population properties for this sample.

\item  This sample has a diverse range of SEDs and SFHs as shown in Figure \ref{fig:sedmodelling_specfits}, covering a range of stellar masses, galaxy types, stellar ages, and SFH shapes. It covers a range of stellar masses from ${\rm 6 < \log M_\star / M_\odot < 12}$, where low mass galaxies have bluer \ug{} colours and higher mass galaxies have a redder, broader range of \ug{} colours, representing the emergence of the Balmer and 4000 \AA{} breaks (see Figure \ref{fig:results-zspec-vs-mass}).

\item The galaxies presented here show a large variation in physical properties, as can be seen in Figure \ref{fig:phys_props_hist}. While the full DJA sample is dominated by low-mass and relatively dust-free galaxies, the massive galaxies in this sample (those with ${\rm \log M_\star / M_\odot > 10}$) tend to have higher SFRs than the bulk sample and span a broader range in sSFRs and \av{}s.

\item Our fitting method is sensitive to a range of star-formation rate timescales, driven by the rich information of the optical and UV stellar continua of the spectra. In Section \ref{subsec:sfrs}, we show the sensitivity of our sample to ${\rm SFR_{10}}$ and ${\rm SFR_{3}}$, where these SFR timescales correlate with the H$\alpha$+[N\,{\sc ii}] EW, a known tracer of recent star-formation on timescales less than 20 Myr.

\item  The typical scatter on the ${\rm M_\star - SFR}$ plane increases from long timescales (${\rm SFR_{100}}$; 1.36 dex) to short (${\rm SFR_{10}}$ and ${\rm SFR_{3}}$; 2.2 and 2.25 dex respectively), which is a reflection of the diverse shapes of the recent SFHs of galaxies. This is further explored in Figure \ref{fig:dja_nirspec_sfr10_sfr100_vs_redshift}, where we find galaxies with a range of rising and falling SFHs within the last 100 Mpc.  When parametrised by ${\rm \log SFR_{10} / SFR_{100}}$, we find young, low-mass galaxies occupy the extremes of recent SFHs shapes, whereas older massive galaxies tend toward flatter recent SFHs. Interestingly, a unique space is carved out by mini-quenched or lulling galaxies, that show the most extreme downturns in recent star-formation.

\item We explore a host of physical properties and their changes across rest-frame \ugis{} colour-colour diagrams and the dust-corrected versions of these diagrams. Massive galaxies range from quiescent to star-forming to dusty. We find a tight sequence of quiescent galaxies within the colour selection box, which are largely dust free and tend to be the oldest galaxies with high $t_{90}$ timescales. Dusty galaxies occupy an extreme version of the colour space corresponding to the reddest \gi{} colours (reddest stellar continua) while showing a larger range in \ug{} colours (ranging from weak to strong Balmer breaks).

\item Galaxies experiencing a recent downturn in star-formation lie just to the bottom left in the quiescent selection box. This follows the evolutionary expectation that as their stellar populations grow older, these galaxies will enter the quiescent region of the diagram as a result of the reddening of their stellar continua and increasingly strong Balmer break.

\item Galaxies with high \av{} ($>1$ mag) occupy the high-mass end of the galaxy main-sequence and lie predominantly within $\pm 1$ dex of the \cite{Popesso2023} SFMS relation. The steepness of the attenuation curve strongly varies with dust attenuation: galaxies with large \av{} have greyer attenuation curves, while galaxies with very low \av{} have steeper curves. We find an interesting correlation with sSFR, where galaxies with the lowest sSFRs have the steepest attenuation curves.

\end{itemize}

Our study provides a new, homogeneously modelled catalogue of SEDs, SFHs and stellar population properties for publicly available PRISM spectra from the Dawn \jwst Archive. We have shown that this catalogue can be applied to a diverse range of studies of various galaxy populations, from the star-formation rates and SFH shapes of UV-bright high-redshift sources to the properties of massive and dusty galaxies. All the data products presented in this paper will be released and made publicly available on the DJA.


\begin{acknowledgements}

The data products presented herein were retrieved from the Dawn \jwst Archive (DJA). DJA is an initiative of the Cosmic Dawn Center (DAWN), which is funded by the Danish National Research Foundation under grant DNRF140.

This work has received funding from the Swiss State Secretariat for Education, Research and Innovation (SERI) under contract number MB22.00072, as well as from the Swiss National Science Foundation (SNSF) through project grants 200020\_207349 and 2000-1-243073.

AdG acknowledges support from a Clay Fellowship awarded by the Smithsonian Astrophysical Observatory. AdG acknowledges funding by the European Union (ERC, FIRST-GIANTS, 101221926). AdG is supported by the Lise Meitner Excellence program of the Max Planck Society. WMB would like to acknowledge support from DARK via the DARK Fellowship. This work was supported by a research grant (VIL54489) from VILLUM FONDEN. L.A.B. acknowledges support from the Dutch Research Council (NWO) under grant VI.Veni.242.055 (\url{https://doi.org/10.61686/LAJVP77714}). AJC gratefully acknowledges support from the Cosmic Dawn Center through the DAWN Fellowship. ORC is supported by National Science Foundation Astronomy \& Astrophysics Postdoctoral Fellowship Award No. 2503202. KEH acknowledges support from the Independent Research Fund Denmark (DFF) under grant 5251-00009B. DN is supported by NASA grants ATP-21-0013, ATP-23-0002, and ADSPS-23-0007. JW gratefully acknowledges support from the Cosmic Dawn Center through the DAWN Fellowship.

The work presented in this paper is based on observations made with the NASA/ESA/CSA \jwst. The data were obtained from the Mikulski Archive for Space Telescopes at the Space Telescope Science Institute, which is operated by the Association of Universities for Research in Astronomy, Inc., under NASA contract NAS 5-03127 for \jwst. These observations are associated with the following programmes: \#1180, \#1181, \#1208 \#1210, \#1211, \#1212, \#1213, \#1214, \#1215, \#1286, \#1287, \#1345, \#1433, \#2198, \#2282, \#2561, \#2565, \#2750, \#2756, \#2767, \#3073, \#3215, \#4106, \#4233, \#4446, \#4568, \#5019, \#5224, \#5545, \#5997, \#6368, \#6541, \#6585, \#8018, \#8060.

{\sl Computational facilities:}
(Part of) The computations were performed at University of Geneva using Baobab/Yggdrasil HPC service.
(Part of) This work was made possible by utilising the CANDIDE cluster at the Institut d’Astrophysique de Paris. The cluster was funded through grants from the PNCG, CNES, DIM-ACAV, the Euclid Consortium, and the Danish National Research Foundation Cosmic Dawn Center (DNRF140). It is maintained by Stephane Rouberol.

{\sl Software:}
We would like to thank the authors of the following software:
{\sc ipython} \citep{Perez2007},
{\sc jupyter} \citep{jupyter},
{\sc astropy} \citep{Astropy2013,Astropy2018},
{\sc matplotlib} \citep{Hunter2007},
{\sc numpy} \citep{numpy},
{\sc scipy} \citep{Virtanen2020},
{\sc eazy-py} \citep{Brammer2008},
{\sc grizli} \citep{Brammer2023},
{\sc msaexp} \citep{brammer_2023_msaexp},
{\sc bagpipes} \citep{Carnall2018, Carnall2019}.

\end{acknowledgements}

\bibliography{bib}{}
\bibliographystyle{aa}

\onecolumn
\begin{appendix}

\section{Tables}

Table \ref{tab:phot-cats} gives an overview of the photometric catalogues used in this work and the number of sources used from each catalogue, drawn from the publicly available DJA database. For a full list of JWST programmes that contribute to a particular DJA image mosaic, we refer the reader to the DJA: \protect\url{https://dawn-cph.github.io/dja/imaging/summary/}.

Table \ref{tab:spec-surveys} gives an overview of the spectroscopic surveys used in this work and the number of {\tt grade=3} NIRSpec/PRISM spectra from each survey, along with the relevant references.

Table \ref{tab:line-wavelengths} is a list of all the lines used in the custom version of \bagpipes{} presented in Section \ref{subsec:line_marginalisation}, along with the line wavelengths and fixed line ratios used from \msaexp{}.

{
\setlength{\LTcapwidth}{\textwidth}
\setlength{\LTleft}{\fill}
\setlength{\LTright}{\fill}

\renewcommand{\arraystretch}{1.5}

\begin{longtable}{l l c c}

\caption{A summary of all photometric catalogues used in this work for NIRCam wide band photometric counterparts to the NIRSpec/PRISM spectra. All catalogues are available on the DJA here: \url{https://dawn-cph.github.io/dja/imaging/v7/}. \emph{N}(filt) is the maximum number of wide band filters available for any given source. \emph{N} is the total number of sources in each photometric catalogue used in this work.\\}
\label{tab:phot-cats}\\
\hline
\hline
Photometric catalogue & Field & \emph{N}(filt) & \emph{N} \\
\hline
\endfirsthead

\hline
\hline
Photometric catalogue & Field & \emph{N}(filt) & \emph{N} \\
\hline
\endhead

\hline
\multicolumn{4}{r}{Continued on next page} \\
\endfoot

\hline
\hline
\endlastfoot

\href{https://dawn-cph.github.io/dja/imaging/v7/?search=abell2744clu-grizli-v7.2-fix_phot.fits}{abell2744clu-grizli-v7.2} & Abell 2744 & 7 & 760 \\
\href{https://dawn-cph.github.io/dja/imaging/v7/?search=abell2744par-grizli-v7.0-fix_phot.fits}{abell2744par-grizli-v7.0} & Abell 2744 & 6 & 4 \\
\href{https://xxdawn-cph.github.io/dja/imaging/v7/?search=abell370-full-grizli-v7.1-fix_phot.fits}{abell370-full-grizli-v7.1} & Abell 370 & 7 & 3 \\
\href{https://dawn-cph.github.io/dja/imaging/v7/?search=abell370-full-grizli-v7.6-fix_phot.fits}{abell370-full-grizli-v7.6} & Abell 370 & 7 & 113 \\
\href{https://dawn-cph.github.io/dja/imaging/v7/?search=ceers-full-grizli-v7.2-fix_phot.fits}{ceers-full-grizli-v7.2} & EGS & 6 & 5 \\
\href{https://dawn-cph.github.io/dja/imaging/v7/?search=ceers-full-grizli-v7.4-fix_phot.fits}{ceers-full-grizli-v7.4} & EGS & 7 & 2143 \\
\href{https://dawn-cph.github.io/dja/imaging/v7/?search=cos-infty-v7.4-fix_phot.fits}{cos-infty-v7.4} & COSMOS & 7 & 14 \\
\href{https://dawn-cph.github.io/dja/imaging/v7/?search=cweb-20250607-grizli-v7.5-fix_phot.fits}{cweb-20250607-grizli-v7.5} & COSMOS & 7 & 1726 \\
\href{https://dawn-cph.github.io/dja/imaging/v7/?search=gdn-grizli-v7.3-fix_phot.fits}{gdn-grizli-v7.3} & GOODS-North & 7 & 5 \\
\href{https://dawn-cph.github.io/dja/imaging/v7/?search=gdn-grizli-v7.4-fix_phot.fits}{gdn-grizli-v7.4} & GOODS-North & 7 & 1642 \\
\href{https://dawn-cph.github.io/dja/imaging/v7/?search=gds-grizli-v7.2-fix_phot.fits}{gds-grizli-v7.2} & GOODS-South & 7 & 1912 \\
\href{https://dawn-cph.github.io/dja/imaging/v7/?search=gds-sw-grizli-v7.0-fix_phot.fits}{gds-sw-grizli-v7.0} & GOODS-South & 7 & 339 \\
\href{https://dawn-cph.github.io/dja/imaging/v7/?search=j112716p4228-grizli-v7.0-fix_phot.fits}{j112716p4228-grizli-v7.0} & & 6 & 3 \\
\href{https://dawn-cph.github.io/dja/imaging/v7/?search=macs0416-clu-grizli-v7.0-fix_phot.fits}{macs0416-clu-grizli-v7.0} & MACS J0416.1-2403 & 7 & 3 \\
\href{https://dawn-cph.github.io/dja/imaging/v7/?search=macs0416-full-grizli-v7.6-fix_phot.fits}{macs0416-full-grizli-v7.6} & MACS J0416.1-2403 & 7 & 100 \\
\href{https://dawn-cph.github.io/dja/imaging/v7/?search=macs0417-full-grizli-v7.6-fix_phot.fits}{macs0417-full-grizli-v7.6} & MACS J0417.5-1154 & 7 & 92 \\
\href{https://dawn-cph.github.io/dja/imaging/v7/?search=macs0647-grizli-v7.0-fix_phot.fits}{macs0647-grizli-v7.0} & MACS0647+7014 & 6 & 49 \\
\href{https://dawn-cph.github.io/dja/imaging/v7/?search=macs1149-clu-grizli-v7.0-fix_phot.fits}{macs1149-clu-grizli-v7.0} & MACS J1149.5+2223 & 7 & 3 \\
\href{https://dawn-cph.github.io/dja/imaging/v7/?search=macs1149-full-grizli-v7.6-fix_phot.fits}{macs1149-full-grizli-v7.6} & MACS J1149.5+2223 & 7 & 129 \\
\href{https://dawn-cph.github.io/dja/imaging/v7/?search=macs1423-full-grizli-v7.2-fix_phot.fits}{macs1423-full-grizli-v7.2} & MACSJ1423.8+2404 & 7 & 2 \\
\href{https://dawn-cph.github.io/dja/imaging/v7/?search=macs1423-full-grizli-v7.6-fix_phot.fits}{macs1423-full-grizli-v7.6} & MACSJ1423.8+2404 & 7 & 53 \\
\href{https://dawn-cph.github.io/dja/imaging/v7/?search=primer-cosmos-east-grizli-v7.0-fix_phot.fits}{primer-cosmos-east-grizli-v7.0} & COSMOS & 7 & 8 \\
\href{https://dawn-cph.github.io/dja/imaging/v7/?search=primer-cosmos-east-grizli-v7.4-fix_phot.fits}{primer-cosmos-east-grizli-v7.4} & COSMOS & 7 & 61 \\
\href{https://dawn-cph.github.io/dja/imaging/v7/?search=primer-cosmos-west-grizli-v7.0-fix_phot.fits}{primer-cosmos-west-grizli-v7.0} & COSMOS & 7 & 14 \\
\href{https://dawn-cph.github.io/dja/imaging/v7/?search=primer-cosmos-west-grizli-v7.4-fix_phot.fits}{primer-cosmos-west-grizli-v7.4} & COSMOS & 7 & 5 \\
\href{https://dawn-cph.github.io/dja/imaging/v7/?search=primer-uds-north-grizli-v7.2-fix_phot.fits}{primer-uds-north-grizli-v7.2} & UDS & 7 & 8 \\
\href{https://dawn-cph.github.io/dja/imaging/v7/?search=primer-uds-south-grizli-v7.2-fix_phot.fits}{primer-uds-south-grizli-v7.2} & UDS & 7 & 10 \\
\href{https://dawn-cph.github.io/dja/imaging/v7/?search=rxj2129-grizli-v7.0-fix_phot.fits}{rxj2129-grizli-v7.0} & & 6 & 17 \\
\href{https://dawn-cph.github.io/dja/imaging/v7/?search=sunrise-grizli-v7.5-fix_phot.fits}{sunrise-grizli-v7.5} & & 7 & 43 \\
\href{https://dawn-cph.github.io/dja/imaging/v7/?search=uds-grizli-v8.0-minerva-v2.2-40mas-fix_phot.fits}{uds-grizli-v8.0-minerva-v2.2-40mas} & UDS & 7 & 2631 \\

\end{longtable}

}

{


\setlength{\LTcapwidth}{\textwidth}
\setlength{\LTleft}{\fill}
\setlength{\LTright}{\fill}

\renewcommand{\arraystretch}{1.5}

\begin{longtable}{l l l c l}
\caption{A summary of all spectroscopic surveys from which the NIRSpec/PRISM spectra used in this work are drawn. Surveys include Early Release Science (ERS), Guaranteed Time Observer (GTO), General Observer (GO) and Director's Discretionary (DD) programmes. A full list of all spectroscopic programmes available on the DJA can be found here: \protect\url{https://dawn-cph.github.io/dja/spectroscopy/nirspec/}.\\}
\label{tab:spec-surveys}\\
\hline
\hline
JWST program & Survey & \emph{N} & Reference \\
\hline
\endfirsthead

\hline
\hline
JWST program & Survey & \emph{N} & Reference \\
\hline
\endhead

\hline
\multicolumn{5}{r}{Continued on next page} \\
\endfoot

\hline
\hline
\endlastfoot

\href{https://www.stsci.edu/jwst-program-info/program/?program=1180}{1180} &
JADES &
493 &
\cite{DEugenio2025}
\\

\href{https://www.stsci.edu/jwst-program-info/program/?program=1181}{1181} &
'' &
839 & ''
\\

\href{https://www.stsci.edu/jwst-program-info/program/?program=1210}{1210} &
'' &
145 & ''
\\

\href{https://www.stsci.edu/jwst-program-info/program/?program=1286}{1286} &
'' &
642 & ''
\\

\href{https://www.stsci.edu/jwst-program-info/program/?program=1287}{1287} &
'' &
161 & ''
\\

\href{https://www.stsci.edu/jwst-program-info/program/?program=1208}{1208} &
CANUCS &
485 &
\cite{Willott2022}
\\

\href{https://www.stsci.edu/jwst-program-info/program/?program=1211}{1211} &
GTO WIDE &
284 &
\cite{Maseda2024}
\\

\href{https://www.stsci.edu/jwst-program-info/program/?program=1212}{1212} &
'' &
219 & ''
\\

\href{https://www.stsci.edu/jwst-program-info/program/?program=1213}{1213} &
'' &
137 & ''
\\

\href{https://www.stsci.edu/jwst-program-info/program/?program=1214}{1214} &
'' &
222 & ''
\\

\href{https://www.stsci.edu/jwst-program-info/program/?program=1215}{1215} &
'' &
209 & ''
\\

\href{https://www.stsci.edu/jwst-program-info/program/?program=1345}{1345} &
CEERS-ERS &
230 &
\cite{Finkelstein2023}
\\

\href{https://www.stsci.edu/jwst-program-info/program/?program=1433}{1433} &
MACS-J0647 &
49 &
\cite{Hsiao2024}
\\

\href{https://www.stsci.edu/jwst-program-info/program/?program=2198}{2198} &
PI: Laia Barrufet
&
98 &
\cite{Barrufet2024}
\\

\href{https://www.stsci.edu/jwst-program-info/program/?program=2282}{2282} &
PI: Dan Coe
&
43 &
\\

\href{https://www.stsci.edu/jwst-program-info/program/?program=2561}{2561} &
UNCOVER &
495 &
\cite{Bezanson2022}
\\

\href{https://www.stsci.edu/jwst-program-info/program/?program=2565}{2565} &
PI: Karl Glazebrook
&
230 &
\cite{Nanayakkara2024b}
\\

\href{https://www.stsci.edu/jwst-program-info/program/?program=2750}{2750} &
PI: Pablo Arrabal Haro
&
73 &
\cite{ArrabalHaro2023_pub}
\\

\href{https://www.stsci.edu/jwst-program-info/program/?program=2756}{2756} &
PI: Wenlei Chen 
&
49 &
\cite{Mascia2024}
\\

\href{https://www.stsci.edu/jwst-program-info/program/?program=2767}{2767} &
PI: Patrick Kelly
&
17 &
\cite{Williams2023_prop}
\\

\href{https://www.stsci.edu/jwst-program-info/program/?program=3073}{3073} &
PI: Marco Castellano
&
220 &
\cite{Castellano2024}
\\

\href{https://www.stsci.edu/jwst-program-info/program/?program=3215}{3215} &
JADES Ultra-Deep &
109 &
\cite{Eisenstein2025}
\\

\href{https://www.stsci.edu/jwst-program-info/program/?program=4106}{4106} &
PI: Erica Nelson &
165 &
\cite{Nelson2023_jwstprop}
\\

\href{https://www.stsci.edu/jwst-program-info/program/?program=4233}{4233} &
RUBIES &
1732 &
\cite{deGraaff2025_RUBIES}
\\

\href{https://www.stsci.edu/jwst-program-info/program/?program=4446}{4446} &
PI: Brenda Frye
&
3 &
\cite{Frye2024}
\\

\href{https://www.stsci.edu/jwst-program-info/program/?program=4568}{4568} &
PI: Justin Pierel
&
13 &
\\

\href{https://www.stsci.edu/jwst-program-info/program/?program=5019}{5019} &
PI: Shiying LU
&
43 &
\\

\href{https://www.stsci.edu/jwst-program-info/program/?program=5224}{5224} &
Mirage or Miracle &
527 &
\cite{Naidu2026}
\\

\href{https://www.stsci.edu/jwst-program-info/program/?program=5545}{5545} &
PI: Laia Barrufet
&
133 &
\cite{Barrufet2024_jwstprop}
\\

\href{https://www.stsci.edu/jwst-program-info/program/?program=5997}{5997} &
PI: Tobias Looser
&
153 &
\\

\href{https://www.stsci.edu/jwst-program-info/program/?program=6368}{6368} &
CAPERS (M. Dickinson) &
2773 &
\\

\href{https://www.stsci.edu/jwst-program-info/program/?program=6541}{6541} &
JADES Transient Survey &
71 &
\cite{DeCoursey2025}
\\

\href{https://www.stsci.edu/jwst-program-info/program/?program=6585}{6585} &
D. Coulter &
150 &
\\

\href{https://www.stsci.edu/jwst-program-info/program/?program=8018}{8018} &
PI: Xiaojing Lin
&
524 &
\\

\href{https://www.stsci.edu/jwst-program-info/program/?program=8060}{8060} &
PI: Eiichi Egami
&
160 &
\\

\end{longtable}

}

{

\setlength{\LTcapwidth}{\textwidth}
\setlength{\LTleft}{\fill}
\setlength{\LTright}{\fill}

\renewcommand{\arraystretch}{1.5}

\begin{longtable}{lll}
\caption{Line quantities and their wavelengths used for the flexible line modelling as described in Section \ref{subsec:line_marginalisation}. Fixed line ratios are used in cases of doublets or blended lines; line ratios are drawn from \msaexp{}.\\}
\label{tab:line-wavelengths}\\
\hline
\hline
Line name & Wavelength (\AA) & Line ratios \\
\hline
\endfirsthead

\hline
\hline
Line name & Wavelength (\AA) & Line ratios \\
\hline
\endhead

\hline
\multicolumn{3}{r}{Continued on next page} \\
\endfoot

\hline
\hline
\endlastfoot

${\rm H\alpha}$+[N\,{\sc ii}] & 6564.61, 6549.86, 6585.27 & 1.0, 0.083, 0.25 \\
${\rm H\beta}$ & 4862.73 & 1.0 \\
${\rm H\gamma}$ & 4341.73 & 1.0 \\
${\rm H\delta}$ & 4102.93 & 1.0 \\
${\rm H\epsilon}$ & 3971.236 & 1.0 \\
${\rm H\zeta}$ & 3890.191 & 1.0 \\
${\rm Pa\alpha}$ & 18756.3 & 1.0 \\
${\rm Pa\beta}$ & 12821.7 & 1.0 \\
${\rm Pa\gamma}$ & 10941.2 & 1.0 \\
${\rm Pa\delta}$ & 10052.2 & 1.0 \\
${\rm He I}$ & 10832.057, 10833.306 & 1.0, 1.0 \\
${\rm He I}$ & 3889.75 & 1.0 \\
${\rm He I}$ & 5877.249 & 1.0 \\
${\rm He II}$ & 1640.4 & 1.0 \\
{[O\,{\sc iii}]}${\lambda\lambda4959,5008}$ & 5008.24, 4960.29 & 2.98, 1.0 \\
{[O\,{\sc iii}]}${\rm \lambda}$ & 4364.436 & 1.0 \\
{[O\,{\sc ii}]}${\rm \lambda\lambda3726,3729}$ & 3727.09, 3729.87 & 1, 1.0 \\
{[O\,{\sc i}]}${\rm \lambda6302}$ & 6302.04, 6365.53 & 1, 0.33 \\
{[S\,{\sc ii}]}${\rm \lambda\lambda6716,6731}$ & 6718.29, 6732.67 & 1.0, 1.0 \\
{[S\,{\sc iii}]}${\rm \lambda9068}$ & 9071.1 & 1 \\
{[S\,{\sc iii}]}${\rm \lambda9531}$ & 9533.2 & 1 \\
{[Ar\,{\sc iii}]}${\rm \lambda7138}$ & 7137.77 & 1.0 \\
{[Ar\,{\sc iii}]}${\rm \lambda7753}$ & 7753.19 & 1.0 \\
{[Mg\,{\sc ii}]}${\rm }$ & 2796.352, 2803.531 & 1.0, 1.0 \\
{[Fe\,{\sc ii}]}${\rm \lambda 12570}$ & 12570.2 & 1.0 \\
{[N\,{\sc v}]}${\rm \lambda1487}$ & 1487.0 & 1.0 \\
{[N\,{\sc v}]}${\rm \lambda1240}$ & 1240.81 & 1.0 \\
{[C\,{\sc iii}]}${\rm \lambda1906}$ & 1906.68 & 1.0 \\
{[C\,{\sc iii}]}${\rm \lambda1908}$ & 1908.73 & 1.0 \\
{[C\,{\sc iv}]}${\rm \lambda1549}$ & 1549.48 & 1.0 \\
{[Ne\,{\sc iii}]}${\rm \lambda3867}$ & 3869.87 & 1.0 \\
{[Ne\,{\sc iii}]}${\rm \lambda3968}$ & 3968.59 & 1.0 \\

\end{longtable}

}

\begin{landscape}
{
\setlength{\LTcapwidth}{\textwidth}
\setlength{\LTleft}{\fill}
\setlength{\LTright}{\fill}

\renewcommand{\arraystretch}{1.5}

\begin{longtable}{lccccccccccc}
\caption{Physical properties of the galaxies shown in Figure \ref{fig:sedmodelling_specfits}. A full table for all $\sim12,000$ sources presented in this paper will be released on Zenodo.}\\
\label{tab:phys_props}\\
\hline
\hline
ID & RA & Dec & $z_{\rm spec}$ & log M$_\star$ & SFR$_{3\,\text{Myr}}$ & SFR$_{10\,\text{Myr}}$ & SFR$_{100\,\text{Myr}}$ & $t_{50}$ & $t_{90}$ & A$_V$ & $\delta$ \\
& ${\rm deg}$ & ${\rm deg}$ &  & M$_\odot$ & M${\rm _\odot/yr}$ & M${\rm _\odot/yr}$ & M${\rm _\odot/yr}$ & ${\rm Gyr}$ & ${\rm Gyr}$ & ${\rm mag}$ &  \\
\hline
\endfirsthead

\hline
\hline
ID & RA & Dec & $z_{\rm spec}$ & log M$_\star$ & SFR$_{3\,\text{Myr}}$ & SFR$_{10\,\text{Myr}}$ & SFR$_{100\,\text{Myr}}$ & $t_{50}$ & $t_{90}$ & A$_V$ & $\delta$ \\
& ${\rm deg}$ & ${\rm deg}$ &  & M$_\odot$ & M${\rm _\odot/yr}$ & M${\rm _\odot/yr}$ & M${\rm _\odot/yr}$ & ${\rm Gyr}$ & ${\rm Gyr}$ & ${\rm mag}$ &  \\
\hline
\endhead

\hline
\multicolumn{12}{r}{Continued on next page} \\
\endfoot

\hline
\hline
\endlastfoot

5224\_277193 & $150.09$ & $2.27$ & $14.47$ & $8.77^{+0.21}_{-0.22}$ & $3.24^{+3.01}_{-1.83}$ & $3.99^{+2.35}_{-1.81}$ & $4.67^{+2.54}_{-1.76}$ & $0.12^{+0.01}_{-0.03}$ & $0.05^{+0.01}_{-0.01}$ & $0.15^{+0.11}_{-0.11}$ & $0.02^{+0.08}_{-0.09}$ \\
1287\_183348 & $53.08$ & $-27.86$ & $14.17$ & $9.38^{+0.11}_{-0.15}$ & $7.96^{+4.26}_{-4.24}$ & $9.88^{+3.40}_{-3.31}$ & $14.68^{+4.57}_{-3.25}$ & $0.14^{+0.01}_{-0.01}$ & $0.06^{+0.01}_{-0.01}$ & $0.12^{+0.04}_{-0.04}$ & $0.05^{+0.08}_{-0.09}$ \\
3073\_22600 & $3.50$ & $-30.32$ & $12.35$ & $8.67^{+0.22}_{-0.33}$ & $38.07^{+9.64}_{-7.23}$ & $18.61^{+2.67}_{-2.84}$ & $4.20^{+2.16}_{-1.98}$ & $0.14^{+0.04}_{-0.07}$ & $0.03^{+0.03}_{-0.03}$ & $0.17^{+0.04}_{-0.04}$ & $0.01^{+0.07}_{-0.08}$ \\
2750\_1 & $214.94$ & $52.94$ & $11.41$ & $8.88^{+0.12}_{-0.15}$ & $2.30^{+1.57}_{-1.13}$ & $2.60^{+1.27}_{-0.85}$ & $3.69^{+1.40}_{-0.99}$ & $0.19^{+0.02}_{-0.05}$ & $0.08^{+0.02}_{-0.02}$ & $0.06^{+0.06}_{-0.04}$ & $0.02^{+0.09}_{-0.10}$ \\
1181\_3991 & $189.11$ & $62.24$ & $10.61$ & $9.06^{+0.07}_{-0.08}$ & $37.43^{+4.88}_{-5.78}$ & $24.89^{+3.07}_{-2.61}$ & $8.02^{+1.36}_{-1.12}$ & $0.21^{+0.02}_{-0.07}$ & $0.07^{+0.02}_{-0.03}$ & $0.07^{+0.02}_{-0.02}$ & $-0.01^{+0.09}_{-0.08}$ \\
3215\_265801 & $53.11$ & $-27.77$ & $9.44$ & $8.39^{+0.08}_{-0.08}$ & $15.18^{+2.62}_{-2.20}$ & $8.31^{+1.28}_{-1.23}$ & $2.06^{+0.34}_{-0.30}$ & $0.22^{+0.05}_{-0.10}$ & $0.05^{+0.04}_{-0.02}$ & $0.29^{+0.05}_{-0.05}$ & $-0.08^{+0.06}_{-0.05}$ \\
6368\_13189 & $34.42$ & $-5.15$ & $8.27$ & $9.36^{+0.07}_{-0.06}$ & $24.26^{+11.98}_{-8.14}$ & $29.46^{+9.56}_{-7.62}$ & $17.85^{+3.71}_{-3.61}$ & $0.22^{+0.11}_{-0.12}$ & $0.06^{+0.05}_{-0.03}$ & $0.37^{+0.08}_{-0.07}$ & $-0.05^{+0.08}_{-0.08}$ \\
4233\_977881 & $34.43$ & $-5.11$ & $7.27$ & $10.31^{+0.06}_{-0.05}$ & $6.11^{+11.79}_{-5.39}$ & $10.90^{+15.91}_{-9.17}$ & $76.77^{+29.64}_{-24.41}$ & $0.28^{+0.12}_{-0.09}$ & $0.11^{+0.04}_{-0.02}$ & $0.77^{+0.14}_{-0.14}$ & $-0.05^{+0.07}_{-0.06}$ \\
1181\_926 & $189.08$ & $62.26$ & $6.76$ & $9.83^{+0.04}_{-0.04}$ & $44.37^{+16.20}_{-15.72}$ & $76.51^{+23.53}_{-17.58}$ & $73.95^{+6.91}_{-10.58}$ & $0.08^{+0.16}_{-0.02}$ & $0.03^{+0.01}_{-0.01}$ & $0.52^{+0.06}_{-0.06}$ & $-0.04^{+0.04}_{-0.05}$ \\
1345\_397 & $214.84$ & $52.88$ & $6.01$ & $8.82^{+0.07}_{-0.05}$ & $30.57^{+4.65}_{-4.69}$ & $28.05^{+5.91}_{-5.51}$ & $6.88^{+0.89}_{-0.93}$ & $0.07^{+0.38}_{-0.05}$ & $0.01^{+0.05}_{-0.01}$ & $0.20^{+0.04}_{-0.04}$ & $0.01^{+0.07}_{-0.07}$ \\
4233\_171974 & $34.21$ & $-5.10$ & $5.21$ & $11.03^{+0.06}_{-0.07}$ & $788.34^{+480.76}_{-378.53}$ & $1223.62^{+545.07}_{-374.05}$ & $1080.60^{+267.34}_{-293.56}$ & $0.16^{+0.46}_{-0.11}$ & $0.03^{+0.06}_{-0.01}$ & $2.41^{+0.18}_{-0.21}$ & $0.19^{+0.03}_{-0.03}$ \\
4233\_75646 & $214.92$ & $52.95$ & $4.90$ & $10.76^{+0.01}_{-0.01}$ & $0.06^{+0.53}_{-0.06}$ & $0.09^{+0.59}_{-0.09}$ & $0.23^{+1.61}_{-0.23}$ & $0.39^{+0.02}_{-0.01}$ & $0.23^{+0.01}_{-0.01}$ & $0.34^{+0.03}_{-0.03}$ & $-0.27^{+0.04}_{-0.05}$ \\
2565\_10459 & $34.34$ & $-5.24$ & $3.97$ & $10.93^{+0.02}_{-0.03}$ & $0.00^{+0.05}_{-0.00}$ & $0.00^{+0.05}_{-0.00}$ & $0.01^{+0.09}_{-0.01}$ & $1.16^{+0.02}_{-0.03}$ & $0.49^{+0.38}_{-0.06}$ & $0.30^{+0.03}_{-0.06}$ & $-0.31^{+0.12}_{-0.08}$ \\
6368\_8495 & $150.17$ & $2.44$ & $3.90$ & $8.84^{+0.05}_{-0.04}$ & $7.82^{+5.72}_{-5.44}$ & $35.07^{+6.94}_{-6.87}$ & $7.85^{+0.89}_{-0.72}$ & $0.02^{+0.09}_{-0.00}$ & $0.01^{+0.00}_{-0.00}$ & $0.35^{+0.05}_{-0.05}$ & $-0.02^{+0.06}_{-0.07}$ \\
1212\_4565 & $53.15$ & $-27.87$ & $3.58$ & $10.60^{+0.06}_{-0.04}$ & $89.77^{+27.87}_{-25.89}$ & $81.23^{+25.56}_{-22.93}$ & $116.27^{+38.95}_{-39.86}$ & $0.77^{+0.29}_{-0.34}$ & $0.22^{+0.12}_{-0.08}$ & $2.26^{+0.11}_{-0.11}$ & $-0.07^{+0.05}_{-0.04}$ \\
1215\_5935 & $34.25$ & $-5.23$ & $3.21$ & $9.93^{+0.03}_{-0.03}$ & $0.00^{+0.00}_{-0.00}$ & $0.00^{+0.01}_{-0.00}$ & $0.00^{+0.01}_{-0.00}$ & $1.45^{+0.02}_{-0.02}$ & $0.96^{+0.05}_{-0.06}$ & $0.12^{+0.07}_{-0.06}$ & $-0.01^{+0.09}_{-0.09}$ \\
5545\_38813 & $--$ & $--$ & $3.01$ & $10.47^{+0.05}_{-0.06}$ & $7.68^{+16.62}_{-6.60}$ & $9.49^{+13.97}_{-7.70}$ & $21.52^{+13.58}_{-11.02}$ & $1.15^{+0.18}_{-0.27}$ & $0.44^{+0.11}_{-0.10}$ & $3.05^{+0.14}_{-0.17}$ & $0.10^{+0.05}_{-0.06}$ \\
4233\_69663 & $214.89$ & $52.92$ & $2.81$ & $9.85^{+0.03}_{-0.03}$ & $0.63^{+1.21}_{-0.55}$ & $0.96^{+1.41}_{-0.78}$ & $7.04^{+4.16}_{-3.22}$ & $0.70^{+0.40}_{-0.27}$ & $0.25^{+0.10}_{-0.06}$ & $0.32^{+0.07}_{-0.09}$ & $-0.07^{+0.09}_{-0.08}$ \\
2198\_5446 & $53.03$ & $-27.85$ & $2.71$ & $10.17^{+0.05}_{-0.06}$ & $5.31^{+9.42}_{-3.63}$ & $9.02^{+8.56}_{-5.78}$ & $19.32^{+7.40}_{-6.96}$ & $1.45^{+0.21}_{-0.22}$ & $0.60^{+0.16}_{-0.16}$ & $2.00^{+0.12}_{-0.15}$ & $0.10^{+0.06}_{-0.07}$ \\
1215\_2069 & $34.42$ & $-5.26$ & $2.54$ & $10.53^{+0.01}_{-0.02}$ & $0.00^{+0.01}_{-0.00}$ & $0.00^{+0.01}_{-0.00}$ & $0.00^{+0.02}_{-0.00}$ & $1.98^{+0.02}_{-0.03}$ & $1.36^{+0.18}_{-0.13}$ & $0.24^{+0.04}_{-0.04}$ & $-0.12^{+0.08}_{-0.09}$ \\

\end{longtable}
}
\end{landscape}

\section{Dust-corrected colours}
\label{appendix:dust_correction}

In this work, we study the dust-corrected colours of galaxies in Section \ref{subsec:colour_colour_diagrams}. Given our use of a flexible attenuation law within the SED modelling framework (as put forth in \cite{Salim2018_dust_law} and implemented in {\sc Bagpipes} \citep{Carnall2018}), we can recover the intrinsic spectrum of each source, and thus the intrinsic colours.

First, we scale the observed-frame flux densities $f_{\lambda, obs}$ by the calibration curve $C(\lambda_{obs})$ (as described in Section \ref{subsec:phot_calibration}) to get the photometrically calibrated observed spectrum: $f^C_{\lambda, obs} = f_{\lambda, obs} \cdot C(\lambda)$. We then convert the observed-frame wavelengths and photometrically calibrated flux densities to the rest frame, via $\lambda_{rest} = \lambda_{obs} / (1+z)$ and $f^C_{\lambda, rest} = f^C_{\lambda, obs} \cdot (1+z)$. We then find the intrinsic spectrum, corrected for dust reddening, via the following:
\begin{equation}
\label{eq:spectrum_intrinsic}
    f_{\lambda,int} = f_{\lambda,rest} \cdot 10^{0.4A_\lambda},
\end{equation}
where $f_{\lambda, int}$ is the intrinsic spectrum as a function of rest-frame wavelengths and $A_\lambda$ is the extinction curve derived from our SED modelling. We follow Equations (2)-(4) and (9) in \cite{Salim2018_dust_law} to derive $A_\lambda$ from the posterior quantities $\delta$ (the power-law deviation from the \citep{Calzetti2000} attenuation curve), $B$ (the strength of the $2175 \text{\AA}$ dust bump) and \av{} (the attenuation in the rest-frame $V$ band.)

Following Equation \ref{eq:spectrum_intrinsic}, we derive the intrinsic dust-corrected spectrum and then interpolate this through rest-frame photometric bands such as the tophat filters introduced in \cite{Danso2023} using the softwares \eazy{} \citep{Brammer2008} and \msaexp{} \citep{brammer_2023_msaexp}. We repeat this process for the calibrated spectrum $f^C_{\lambda, obs}$. In doing so, we derive both the rest-frame colours from the photometrically calibrated spectrum (used in the top panels of Figure \ref{fig:ugi_sed_cbar}) and the intrinsic dust-corrected spectrum (used in the lower panels of Figure \ref{fig:ugi_sed_cbar}).

\section{Stellar masses from photometric vs. spectrophotometric SED fitting}
\label{appendix:stellar_mass_phot_vs_spec}

\begin{figure*}[htb!]
\includegraphics[width=0.8\linewidth]{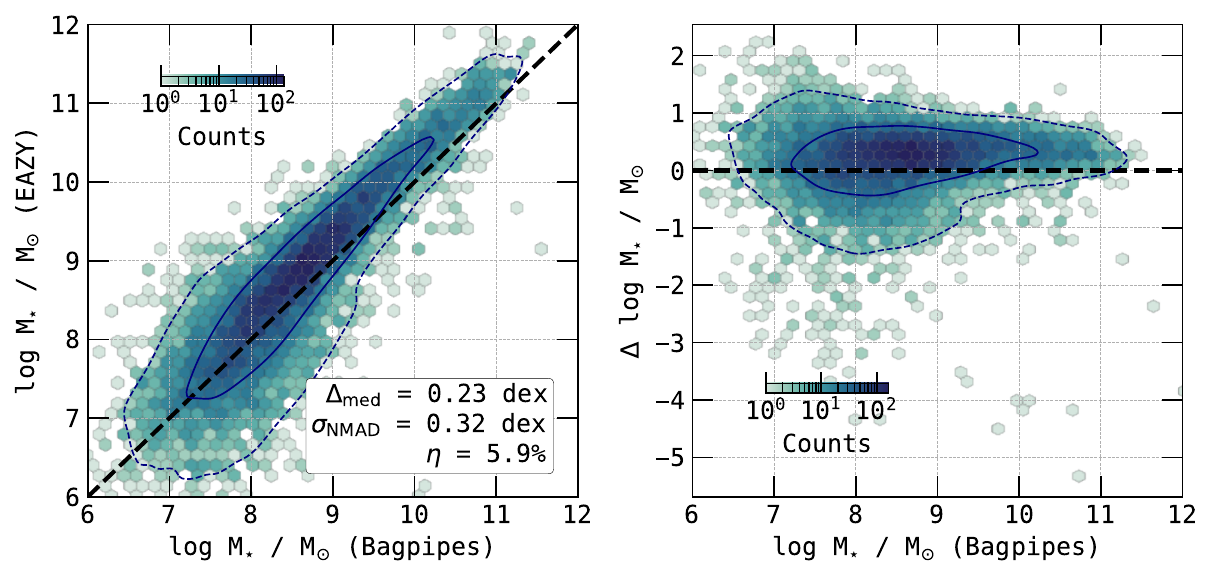}
\caption{Left: A comparison of stellar masses as derived from {\sc Bagpipes} with spectrophotometric fitting and {\sc \eazy{}} with photometry-only fits. Right: ${\rm \Delta \log M_\star / M_\odot}$ as a function of stellar mass from the  {\sc Bagpipes} spectrophotometric fitting. As seen from the left hand plot, there is significant scatter between photometric and spectrophotometric stellar masses, with offset of ${\rm \Delta_\text{med} = 0.23\ dex}$. We also see that a majority of the 1$\sigma$ contour lies above the 1:1 line implying higher masses with \eazy{} photometric fits.}
\label{fig:mstar_bagpipes_vs_eazy}
\end{figure*}

We compare stellar masses derived from photometric SED fitting using the code \eazy{} (available on the DJA, see Table \ref{tab:phot-cats} and links therein) to stellar masses derived from spectrophotometric SED fitting using the catalogue presented in this paper, based on fits with the code \bagpipes{}. The \eazy{} fits utilise all available broad-band photometry from \hst{}/ACS \hst{}/WFC3, and \jwst{}/NIRCam. The spectrophotometric fits utilise broad-band photometry from \jwst{}/NIRCam and spectra from \jwst{}/NIRSpec in the PRISM/CLEAR mode. Figure \ref{fig:mstar_bagpipes_vs_eazy} shows a comparison between the stellar masses derived using the two fitting methods.

While they are in overall agreement, stellar masses derived with \eazy{} tend to be higher, with a median offset of 0.23 dex. The distribution shows a large scatter of 0.32 dex.

\section{H\texorpdfstring{$\alpha$}{Lg} line fluxes and EW}
\label{appendix:ha_flux_eqw}

As described in Section \ref{subsec:djapipes-eval}, the custom line fitting functionality developed for this project and implemented in \bagpipes{} returns measured line fluxes for a host of emission lines (see Table \ref{tab:line-wavelengths} for a full list of added line components). Here, we compare the line fluxes and EW of the ${\rm H\alpha}$+{[N\,{\sc ii}]} line complex derived from the default and flexible \bagpipes{} fitting methods to the line fluxes measured using the \msaexp{} software (available on the DJA).

Given the resolution of the PRISM/CLEAR mode and the often blended nature of the ${\rm H\alpha}$ and {[N\,{\sc ii}]} emission lines, for the purpose of our study we calculate a combined ${\rm H\alpha}$+{[N\,{\sc ii}]} line flux and EW. For the flexible fitting method, the final line fluxes are combined from two underlying measurements: (a) the ${\rm H\alpha}$ and {[N\,{\sc ii}]} line fluxes associated with the \bagpipes{} model, and (b) the ${\rm H\alpha}$ and {[N\,{\sc ii}]} flues measured from the flexible Gaussian components added on-the-fly during the fitting procedure. The fluxes from these methods are added and then down-scaled by the calibration polynomial, such that the fluxes can be compared to those derived from \msaexp{} (which are measured from the observed spectrum). For the default fitting method, in the absence of the additional flexible line components, the combined ${\rm H\alpha}$+{[N\,{\sc ii}]} line flux is found by adding the ${\rm H\alpha}$ and {[N\,{\sc ii}]} fluxes associated with the \bagpipes{} model and downscaling this sum by the calibration polynomial.

\begin{figure*}[htb!]
\includegraphics[width=0.85\linewidth]{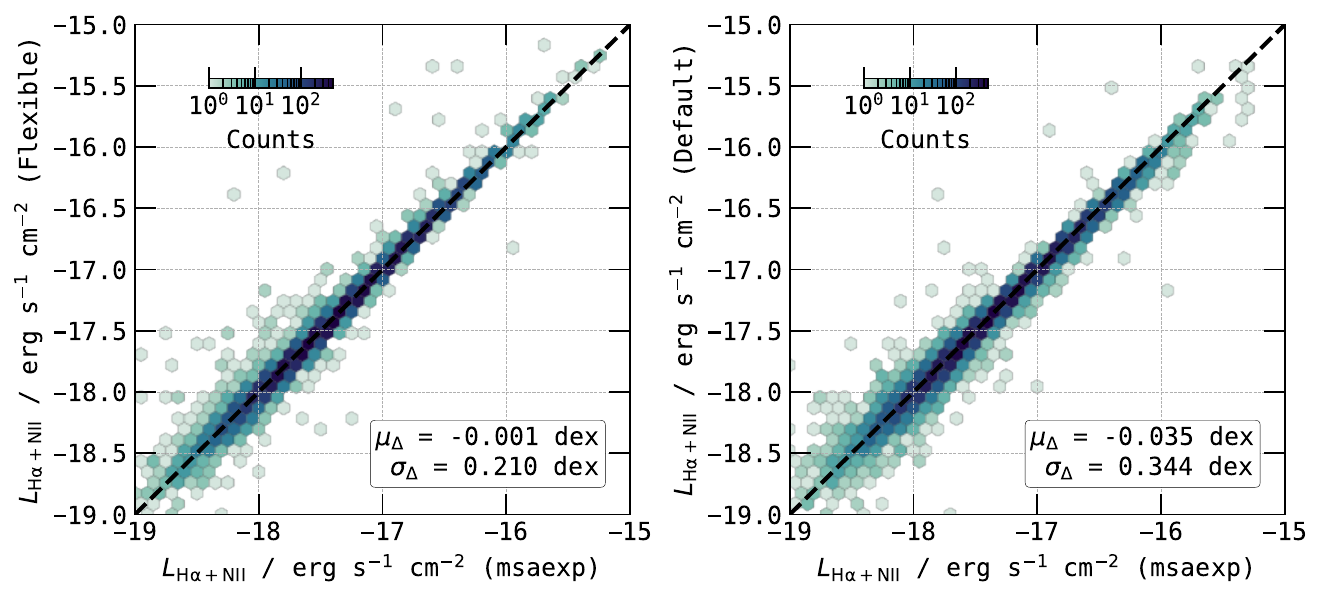}
\caption{Left: ${\rm H_\alpha}$+{[N\,{\sc ii}]} line fluxes from \msaexp{} compared with the linefluxes from our modified version of Bagpipes including the flexible line modelling. Right: same comparison with the default Bagpipes fits. As evidenced from the average scatter between the two methods, the flexible line modelling (${\rm \sigma_\Delta=0.21\ dex}$) yields a better agreement with line fluxes from \msaexp{} than the default Bagpipes modelling (${\rm \sigma_\Delta=0.34\ dex}$).}
\label{fig:ha_nii_flux_comparison_flex_vs_def}
\end{figure*}

The results of the two methods are shown in Figure \ref{fig:ha_nii_flux_comparison_flex_vs_def}. For both methods, the ${\rm H\alpha}$+{[N\,{\sc ii}]} line fluxes are in agreement with the \msaexp{} measurements. The flexible line fitting method is in closer agreement (with a mean offset of -0.001 dex and an average scatter of 0.21 dex) than the default fitting method (with a mean offset of -0.035 dex and an average scatter of 0.344 dex). As discussed in Section \ref{subsec:djapipes-eval}, this improvement is by design, as the flexible line fitting method effectively captures any residual flux that the \bagpipes{} model cannot reproduce.

The EW of the ${\rm H\alpha}$+{[N\,{\sc ii}]} line complex is calculated as the total line flux of the complex divided by the flux density of the underlying stellar continuum at 6564.61 \AA, the wavelength of the ${\rm H\alpha}$ used in \msaexp{} (see Table \ref{tab:line-wavelengths}). As the ${\rm H\alpha}$+{[N\,{\sc ii}]} line flux calculated above in associated with the down-scaled SED fit, the continuum is also estimated from the SED fit down-scaled by the calibration polynomial. On this down-scaled SED fit, the stellar continuum is evaluated at 6564.61 \AA by fitting a first-order polynomial to the flux density in the following two windows flanking the line complex: $6000-6400$ \AA and $67-7100$ \AA.

\begin{figure*}[htb!]
\includegraphics[width=0.75\linewidth]{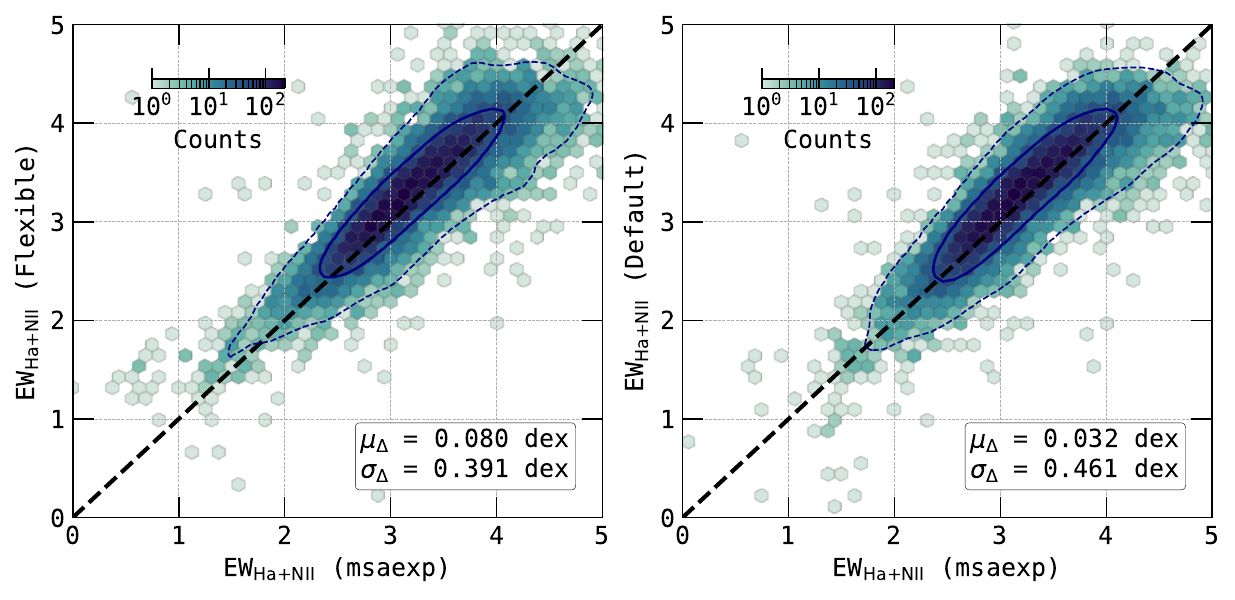}
\caption{Left: ${\rm H\alpha}$+{[N\,{\sc ii}]} EW from \msaexp{} compared with the EW from our modified version of Bagpipes including the flexible line modelling. Right: same comparison with the default Bagpipes fits. As evidenced from the average scatter between the two methods, the flexible line modelling (${\rm \sigma_\Delta=0.39\ dex}$) yields a better agreement with line fluxes from \msaexp{} than the default Bagpipes modelling (${\rm \sigma_\Delta=0.46\ dex}$). We note that in both cases, the scatter is significant.}
\label{fig:ha_nii_eqw_comparison_flex_vs_def}
\end{figure*}

The ${\rm H\alpha}$+{[N\,{\sc ii}]} EW from the flexible and default methods, in comparison with those from \msaexp{}, are shown in Figure \ref{fig:ha_nii_eqw_comparison_flex_vs_def}. Both methods scale with the ${\rm H\alpha}$+{[N\,{\sc ii}]} EW measurements from \msaexp{}. The mean offset is similarly small in both cases, while the flexible method yields a smaller scatter (${\rm \sigma_\Delta=0.39\ dex}$) than the default method (${\rm \sigma_\Delta=0.46\ dex}$).

\section{Colour-colour diagrams: comparison to previous selections}
\label{appendix:uvj_ugi_selections}

\begin{figure*}[htb!]
\includegraphics[width=0.5\linewidth]{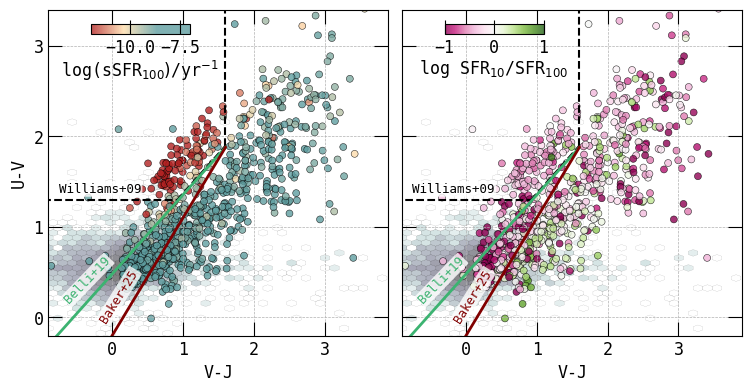}
\includegraphics[width=0.5\linewidth]{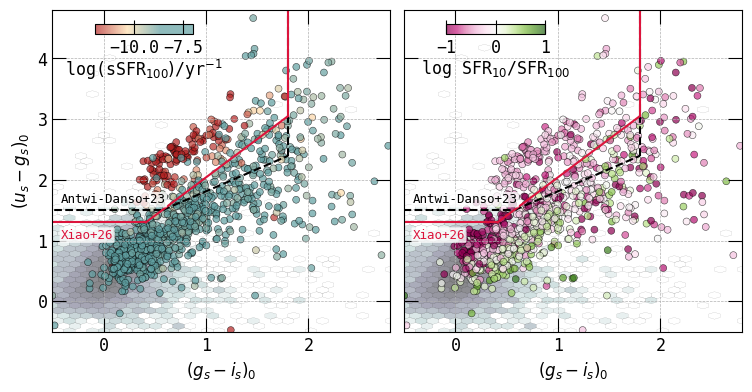}
\caption{Top row: $UVJ$ colour-colour diagrams with overlaid selections from \cite{Williams2009, Belli2019, Baker2024_quiescent_SFHs}. Bottom row: \ugis{} colour-colour diagrams as in Figure \ref{fig:ugi_sed_cbar}, with overlaid selections from \cite{Danso2023} and \citep{Xiao2026}.}
\label{fig:dja_nirspec_observed_UVJ_selections}
\end{figure*}

Colour selections have long been used to efficiently separate quiescent and star-forming galaxies, with the goal being to optimise the completeness of the quiescent galaxy sample while reducing contaminants to the extent possible. In Figure \ref{fig:dja_nirspec_observed_UVJ_selections}, we show the rest-frame $UVJ$ and \ugis{} colour-colour diagrams, and common colour-selections from the literature.

The top two panels in this figure show the $UVJ$ diagram coloured by sSFR on the left and ${\rm \log SFR_{10}/SFR_{100}}$ to the right; over-plotted are the selections from \cite{Williams2009}, \cite{Belli2019} and \cite{Baker2024_quiescent_SFHs}. While the selection from \cite{Williams2009} identifies galaxies with the lowest sSFRs, it misses some galaxies that scatter towards bluer U-V colours. The selections from \cite{Belli2019} and \cite{Baker2024_quiescent_SFHs} capture these galaxies as well as galaxies with low ${\rm \log SFR_{10}/SFR_{100}}$ (which could be recently quenched galaxies or galaxies with a recent downturn in star-formation -- but not yet fully quenched). Qualitatively, the selection from \cite{Williams2009} will have the lowest completeness but least contamination, while \cite{Baker2024_quiescent_SFHs} will have the highest completeness but also highest contamination from blue, star-forming galaxies.

The bottom two panels in Figure \ref{fig:dja_nirspec_observed_UVJ_selections} show the \ugis{} diagram with the same colour schemes as in the $UVJ$ diagram above it. Over-plotted is the original \ugis{} colour selection from \cite{Danso2023} and a recently proposed modification from \cite{Xiao2026}, based on a study of massive galaxies with spectrophotometric fits drawn from this paper. The selection from \cite{Danso2023} encloses a majority of the galaxies with low sSFRs in this sample. It also includes star-forming galaxies with high sSFRs at the boundary of the selection box, as also discussed within that study and in \cite{Zhang2025}. The selection in \cite{Danso2023} therefore has high completeness but also higher contamination rates towards higher redshifts \citep[see discussion in][]{Danso2023}.

\cite{Xiao2026} proposed a selection which extends the \ugis{} to bluer \ug{} colours to capture recently quenched systems while reducing the number of star-forming galaxy contaminants. From Figure \ref{fig:dja_nirspec_observed_UVJ_selections}, we see that this selection captures more galaxies with low ${\rm \log SFR_{10}/SFR_{100}}$ than the traditional \ugis{} selection, but still misses the bulk of sources with low ${\rm \log SFR_{10}/SFR_{100}}$. Importantly this new selection reduces the number of star-forming galaxy interlopers within the \ugis{} selection from \citep{Danso2023}, but increases the number of lower mass galaxies that fall into the selection box as a result of including bluer \ug{} colours. The exact completeness and contamination rates of this selection are yet to be explored with an unbiased sample.

\end{appendix}

\end{document}